\documentclass[fleqn,usenatbib]{mnras}

\usepackage{newtxtext,newtxmath}
\usepackage{xspace,color}
\usepackage[T1]{fontenc}
\usepackage{ae,aecompl}
\newcommand{\azul}[1]{{{\color{blue} #1}}\xspace}

\usepackage{graphicx}	
\usepackage{amsmath}	
\usepackage{comment}    
\usepackage{float}
\usepackage{epstopdf}
\usepackage[colorinlistoftodos]{todonotes}  %
\usepackage{soul}

\newcommand{\angstrom}{\textup{\AA}}

\title[sulphur  abundance estimates]{Chemical abundances in Seyfert galaxies -- X. Sulphur  abundance estimates}
\author[O. L. Dors]{
Oli L. Dors$^{1}$\thanks{E-mail: olidors@univap.br}, M. Valerdi$^{2}$\thanks{E-mail: mabel.astro@gmail.com},
R.~A. Riffel$^{3}$, R. Riffel$^{4}$, M. V. Cardaci$^{5,6}$,
G.~F. H\"agele$^{5,6}$, Mark Armah$^{4}$, \newauthor
~M. Revalski$^{7}$, S.~R. Flury$^{8}$, P. Freitas-Lemes$^{1}$,
E.~B.~Am\^{o}res$^9$, A.~C. Krabbe$^{1}$, 
L. Binette$^{10}$, A. Feltre$^{12}$, \newauthor
~T. Storchi-Bergmann$^{4}$
\\
$^{1}$Universidade do Vale do Para\'iba, Av. Shishima Hifumi, 2911, Cep
12244-000, S\~ao Jos\'e dos Campos, SP, Brazil \\
$^{2}$Instituto Nacional de Astrof{\'i}sica, \'Optica y Electr\'onica (INAOE), Luis E. Erro No. 1, Sta. Ma. Tonantzintla, Puebla, C.P. 72840, M\'exico. \\
$^{3}$Departamento de F\'isica, Centro de Ci\^encias Naturais e Exatas, Universidade Federal de Santa Maria, 97105-900, Santa Maria, RS, Brazil\\
$^{4}$ Departamento de Astronomia, Universidade Federal do Rio Grande do Sul, Av. Bento Gon\c calves 9500, Porto Alegre, RS, Brazil\\
$^5$ Facultad de Ciencias Astron\'omicas y Geof\'{\i}sicas, Universidad Nacional de La Plata, Paseo del Bosque s/n, 1900 La Plata, Argentina.\\
$^6$ Instituto de Astrof´ısica de La Plata (CONICET-UNLP), La Plata, Avenida Centenario (Paseo del Bosque) S/N, B1900FWA, Argentina\\
$^7$ Space Telescope Science Institute, 3700 San Martin Drive, Baltimore, MD 21218, USA \\
$^{8}$ Department of Astronomy, University of Massachusetts Amherst, Amherst, MA 01002, USA \\
$^9$ Departamento de F{\'i}sica, Universidade Estadual de Feira de Santana, Av. Transnordestina, S/N, CEP 44036-900 Feira de Santana, BA, Brazil \\
$^{10}$ Instituto de Astronom\'{\i}a, Universidad Nacional Aut\'onoma de M\'exico, A.P. 70-264, 04510 M\'exico, D.F., M\'exico, M\'exico \\
$^{11}$ D\'epartement de physique, de g\'enie physique et d'optique,  Universit\'e Laval, Qu\'ebec, QC G1V 0A6, Canada \\
$^{12}$ Osservatorio di Astrofisica e Scienza dello Spazio di Bologna, Via P. Gobetti 93/3, 40129 Bologna, Italy
}

\date{Accepted XXX. Received YYY; in original form ZZZ}

\pubyear{2022}

\begin{document}
\label{firstpage}
\pagerange{\pageref{firstpage}--\pageref{lastpage}}
\maketitle

\begin{abstract}
For the first time, the sulphur  abundance relative to hydrogen  (S/H) in the Narrow Line Regions of a sample of Seyfert~2 nuclei (Sy~2s) has been derived via  direct estimation of the electron temperature. Narrow  emission line intensities from the SDSS DR17 [in the wavelength range $3000<\lambda($\AA$)<9100$] and from the literature for a sample of 45 nearby ($z\:<0.08$)  Sy~2s were considered. Our direct estimates indicate that Sy~2s have similar temperatures in the
gas region  where most of the $\rm S\rm ^{+}$ ions are located in comparison with that of star-forming regions (SFs). However, Sy~2s present higher temperature values  ($\sim 10\,000$ K) in the region 
where most of the $\rm S\rm ^{2+}$  ions are located relative to that of SFs.   We derive  the total sulphur  abundance in the range of $6.2 \: \la 12+\log(\rm S/H) \: \la \: 7.5$, corresponding to 0.1  - 1.8 times the solar value. 
These sulphur  abundance values  are lower by $\sim 0.4$ dex than those derived in SFs with similar metallicity, indicating a distinct chemical enrichment of the ISM for these object classes. The  S/O values
 for our  Sy~2  sample present an abrupt ($\sim 0.5$ dex)
 decrease with increasing O/H for the high metallicity regime [$\rm 12+\log(O/H) \: \ga 8.7)$], what is not seen for the SFs. However, when our Sy~2 estimates are combined with those from a  large sample of star-forming regions, we did not find any dependence between S/O and O/H.
\end{abstract}




\section{Introduction}
Sulphur  is mainly produced via $\alpha$-capture in the inner layers of massive stars (e.g. \citealt{1995ApJS..101..181W, 2013ARA&A..51..457N}) and it is  a truly non-refractory element in the interstellar medium (ISM). Due to these features, 
the sulphur  abundance and its abundance relation with the oxygen (S/O) place constraints on stellar nucleosynthesis calculations,  variations of the Initial Mass Function (IMF) of stars as well as in the analysis of the oxygen depletion onto dust grains  (e.g. \citealt{1989ApJ...345..282G, 1996ARA&A..34..279S, 2004AJ....127.2284H}).

Over time, several  studies have obtained
sulphur  (and other $\alpha$-elements) and oxygen abundances in star-forming regions (\ion{H}{ii} regions and \ion{H}{ii} galaxies, hereafter SFs; e.g. \citealt{1978MNRAS.183P...1P, 1978ApJ...222..821S, 1988MNRAS.235..633V, 1997A&A...322...41C, 1997ApJ...489...63G, 2002A&A...391.1081V, 2003ApJ...591..801K, 2006A&A...449..193P, 2008MNRAS.383..209H, 2009A&A...508..615L, 2013ApJ...775..128B, 2016MNRAS.456.4407D, 2019MNRAS.487.3221F, 2020MNRAS.496.1051A, 2020ApJ...893...96B, 2021ApJ...915...21R}). However, the S/O versus  O/H relation  is still ill-defined. In fact, some authors (e.g. \citealt{1988MNRAS.235..633V, 1991MNRAS.253..245D, 2016MNRAS.456.4407D, 2022MNRAS.511.4377D}) have found evidence that S/O decreases as 
O/H increases. On the other hand, constant S/O abundance over a wide range of O/H (a gas phase metallicity tracer\footnote{For a review see \citet{2019A&ARv..27....3M} and \citet{2019ARA&A..57..511K}.}) is supported by a growing body of studies 
(e.g. \citealt{1989ApJ...345..282G, 2003ApJ...591..801K, 2006A&A...448..955I, 2011A&A...529A.149G, 2020ApJ...893...96B, 2021ApJ...915...21R}). 

Abundance estimates in stellar atmospheres, derived from absorption features, 
have confirmed the above  contradictory
results and several scenarios have been reported:  ($i$) a constant increase of the S/Fe abundance ratio as metallicity\footnote{The metallicity in stars is usually traced by Fe/H abundance ratio (e.g., \citealt{2004A&A...420..183A}).} decreases (e.g., \citealt{2001ApJ...557L..43I, 2002ApJ...573..614T}), ($ii$) an increase of S/Fe followed by a constant value at the metal-poor regime
as metallicity decreases (\citealt{2004A&A...415..993N, 2007A&A...469..319N}) and ($iii$) a bimodal behaviour of S/Fe at the metal-poor regime \citep{2005A&A...441..533C}. However, recent chemical abundance  determinations in stellar atmospheres  have found
a decrease of S/Fe with  increasing Fe/H \citep{2020A&A...634A.136C, 2022A&A...657A..29L}. Interestingly, abundance estimates based on
absorption lines in damped Ly$\alpha$ (DLA) systems \citep{2000ApJ...536..540C} showed a decrease of S/Zn with the increase of Zn/H (a metallicity tracer, \citealt{1997ApJ...486..665P}),
indicating that $\alpha$-element burning happens at different times for different elements in massive stars (see also \citealt{2001MNRAS.325..767B, 2002ApJ...566...68P, 2013MNRAS.435.1727F, 2014A&A...572A.102F}).
However, gas-phase abundances in DLAs must be corrected for dust depletion effects, producing additional difficulties in the interpretation of abundance ratio trends (e.g. \citealt{2022ApJ...935..105R}). 

Sulphur  and oxygen abundances have also been largely derived
for Planetary Nebulae (PNe, e.g. \citealt{1980ApJ...240...99B, 1983ApJS...51..211A, 2004A&A...423..199C, 2008ApJ...672..274B, 2017MNRAS.468..272C, 2018A&A...615A..29P, 2018A&A...620A.169W, 2021MNRAS.508.2668E, 2022MNRAS.510.5444G}).
In particular, \citet{2018ApJ...853...50F}, who combined S and O abundances, obtained a clear decrease of S/O with O/H for 10 PNe in the Andromeda Galaxy (M\,31) with estimates relying on data from the literature. Additionally, these authors  found that their sample of PNe have abundance estimates $\sim$0.2–0.4 dex lower than the expected  sulphur -to-oxygen  abundance solar value assuming $\rm \log(S/O)_{\odot}=-1.43$ \citep{1998SSRv...85..161G, 2001ApJ...556L..63A}. 
This discrepancy has previously been attributed to the inadequacy of the Ionization Correction Factors (ICFs) used to correct the presence of unobserved sulphur  ions 
(the so-called “sulphur  anomaly”) by \citet{2004AJ....127.2284H} and  \citet{2010ApJ...711..619M}. 
However,
the PN abundance estimates by \citet{2018ApJ...853...50F} are in
agreement with those derived in \ion{H}{ii} regions also located at the Andromeda Galaxy by \citet{2012MNRAS.427.1463Z},
confirming their results. Moreover, \citet{2014MNRAS.440..536D}, who computed a large grid of photoionization models that covers a wide range of physical
parameters and is representative of most  observed PNe, proposed a robust ICF for the sulphur  and, by using optical observational data for a large sample, confirmed that S/O decreases with O/H.
However, it is worth to mention that, contrary to presently accepted thinking, 
\citet{2009ApJ...700.1299J} showed that sulphur  can deplete by up to $\sim$1 dex, which might account for some of the decrease observed.

Contrary to SFs, stars, DLA systems and PNe, the sulphur  abundance is poorly known in Active Galactic Nuclei (AGNs)
or only qualitative estimates are available in the literature.  The first (qualitative) sulphur  estimates for this class of objects seems to have been performed by \citet{thaisa90}, who compared the intensity of the  [\ion{N}{ii}]($\lambda \lambda$6548, 6584)/H$\alpha$ and [\ion{S}{ii}]($\lambda \lambda$6716, 31)/H$\alpha$ line ratios predicted by photoionization models with observational data from a sample of 177 Seyfert~2   galaxies. These authors found that
models assuming sulphur  abundances ranging from half to five times the solar abundance reproduce the observational
data.  These estimates can be somewhat uncertain because  
the  model fittings by \citet{thaisa90} do not consider the lines emitted by $\rm S^{2+}$, which  can be the most abundant sulphur  ion and occurs as a result of high ionization degree of the AGNs (e.g. \citealt{2014MNRAS.437.2376R, 2021MNRAS.505.4289P}).

Recently, \citet{2020MNRAS.496.3209D} proposed a new methodology
for the $T_{\rm e}$-method   -- a   conventional and reliable method \citep{2003A&A...399.1003P, 2017MNRAS.467.3759T}  based on direct estimates of the electron temperature (for a review see \citealt{2017PASP..129h2001P, 2017PASP..129d3001P}) -- which  makes it possible to estimate the O/H abundance in 
Seyfert~2 nuclei (hereafter Sy~2s). Further studies based on this methodology, for the first time, permitted direct abundance estimates of the
argon \citep{2021MNRAS.508.3023M}, neon \citep{2021MNRAS.508..371A} and helium \citep{2022MNRAS.514.5506D} in the narrow line regions (NLRs) of a large sample of Sy~2s. Generally, this class of  AGN presents solar or oversolar metallicities
($\rm 12+\log(O/H) \: \gtrsim \: 8.7$, e.g. \citealt{1978ApJ...222..821S, 2006MNRAS.371.1559G,2020MNRAS.492..468D})
and gas with high ionization.  These features make it possible to measure some auroral lines (e.g. [\ion{O}{iii}]$\lambda4363$, [\ion{N}{ii}]$\lambda5755$) in the high metallicity regime, which are difficult to detect in SFs (e.g. \citealt{1998AJ....116.2805V, 2007MNRAS.382..251D, 2008A&A...482...59D}). However, it is worthwhile to point out some difficulties in applying the $T_{\rm e}$-method to derive abundances in AGNs, for instance: (i) due to the large width of H$\gamma$, in several cases, this Balmer line is blended with the temperature-sensitive auroral line [\ion{O}{iii}]$\lambda4363$ in AGN spectra; (ii) although AGNs have a  high ionization degree, its  high metallicity (e.g. \citealt{2006MNRAS.371.1559G}) 
produces measurements of [\ion{O}{iii}]$\lambda4363$
with  low signal/noise ratio,   which translates into a large abundance uncertainty (e.g. \citealt{2022MNRAS.514.5506D}) and (iii) temperature estimates from distinct gas regions in AGNs are barely found in the literature, making it difficult to carry out any statistical study.
In any case, with the current observational data and methodologies available in the literature
it is possible to obtain (relatively) precise sulphur  and oxygen abundances, producing important constraints to the
studies of stellar nucleosynthesis in the high metallicity regime.
In fact, even the recent stellar nucleosynthesis  models (e.g. \citealt{2018MNRAS.480..538R}) do not consider oversolar metallicities despite  metallicity has an impact on the stellar product (e.g. \citealt{2021A&A...656A..94G}).

Taking  advantage of the availability of spectroscopic data of Sy~2s in the literature, data provided by the Sloan Digital Sky Survey \citep{2000AJ....120.1579Y} and motivated by 
the new methodology proposed by \citet{2020MNRAS.496.3209D}, in this work, the last in a series of ten papers, we present direct S and O abundance estimates for the NLRs of a sample of 45 Sy~2s. This study is organized as follows. In Section~\ref{observ} the observational data
is presented. The methodology used to estimate the sulphur  and oxygen abundances is presented in Sect.~\ref{abund}. The results and discussion are given in Sect.~\ref{rdisc}.
Finally, the key   findings are summarised in Sect.~\ref{conc}.

\section{Observational data} 
\label{observ}

\begin{figure}
\includegraphics[angle=0.0,width=0.47\textwidth]{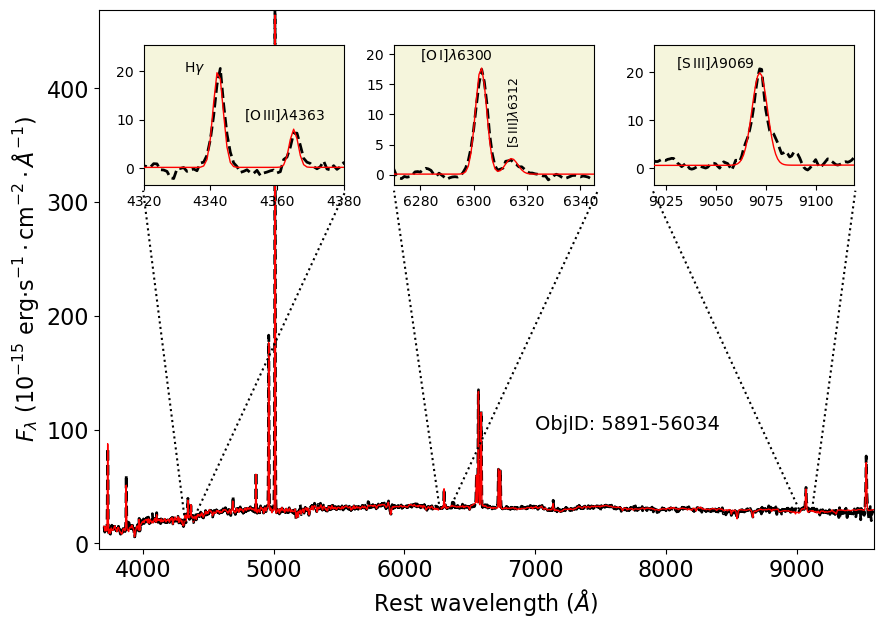}
\caption{Optical spectrum (black colour) for one of the Seyfert~2 nuclei in our sample (see Sect.~\ref{observ}) obtained from the SDSS DR17. The fitting to the emission-line profiles using the {\sc ifscube} code is represented in red colour. The measured emission-lines and their corresponding wavelengths are indicated. Boxes show enlargements of regions around some weak lines, as indicated.}
\label{fig1}
\end{figure}

\begin{figure}
\includegraphics[angle=0.0,width=0.87\textwidth]{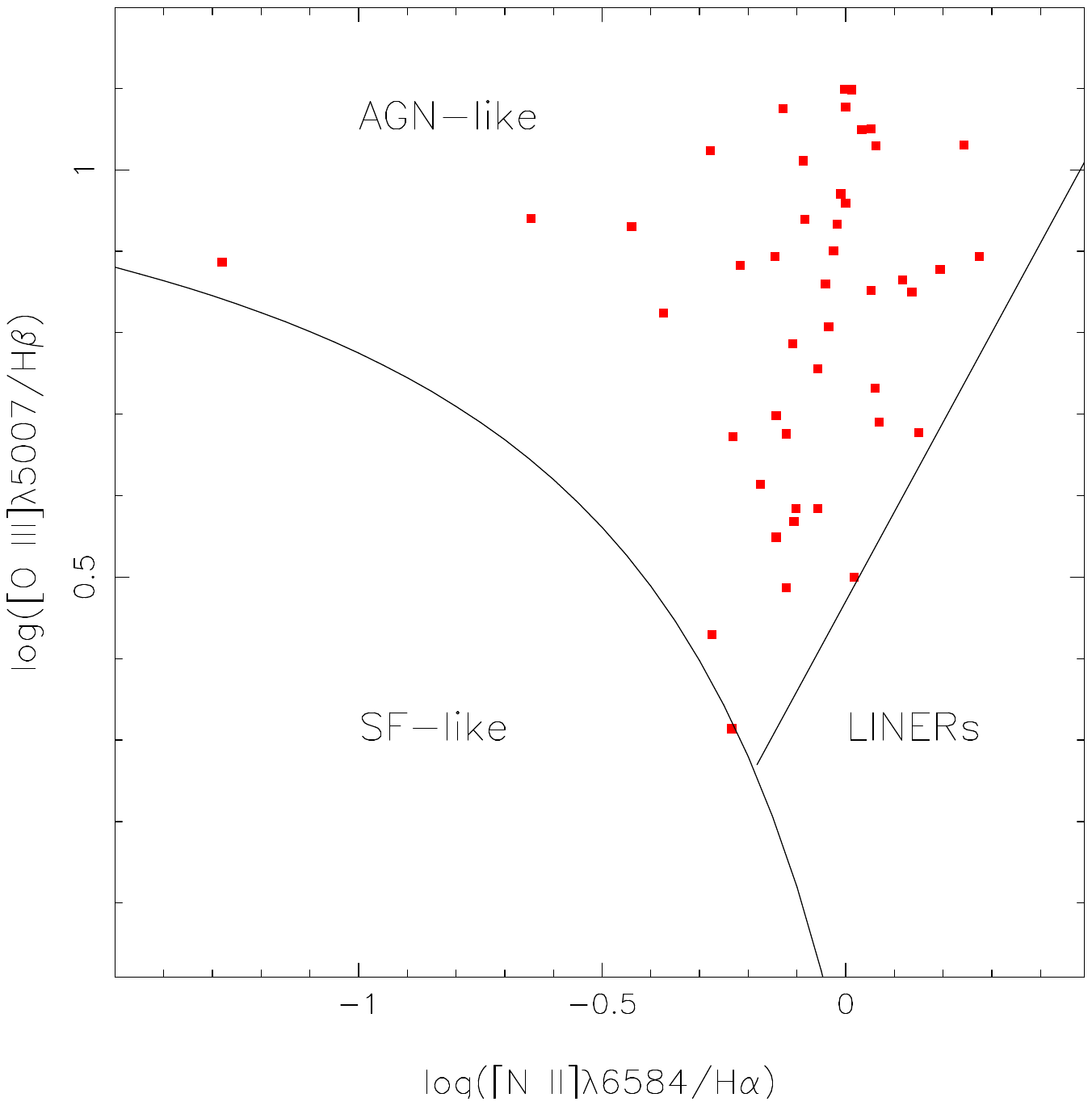}
\caption{log([\ion{O}{iii}]$\lambda 5007$/H$\beta$) vs.\ log([\ion{N}{ii}]$\lambda 6584$/H$\alpha$]) diagnostic diagram. Red points represent the Sy~2 nuclei in our sample (see Sect.~\ref{observ}).  
The black solid curve represents  the criterion (Eq.~\ref{eq1}) proposed by \citet{2001ApJ...556..121K} to separate AGN-like and SF-like objects. The black solid line represents the criterion (Eq.~\ref{eq2}) proposed by \citet{2010MNRAS.403.1036C} to separate AGN-like objects and LINERs.}
\label{fig2}
\end{figure}

In order to obtain a sample of Sy~2s  with observational intensities
of narrow (Full Width Half Maximum  $< \: 1000\: \rm km\: s^{-1}$) optical emission lines, we used  spectroscopic data made available through the Sloan Digital Sky Survey Data Release 17\footnote{SDSS DR17 spectroscopic data are available at \url{https://dr17.sdss.org/optical/plate/search}}, SDSS DR17 \citep{2022ApJS..259...35A}. 
In the present study, the procedures for selection of Sy~2s, emission line measurements, reddening correction and the
stellar population continuum subtraction 
 were the same as  described by \citet{2022MNRAS.514.5506D}, therefore, we summarized these processes below. 

For each spectrum downloaded from the SDSS DR17,
we performed the extinction correction using the 
\citet{1989ApJ...345..245C} law assuming the parameterized extinction coefficient  $R_{V} = 3.1$. Thereafter, the stellar population continuum was subtracted from the spectra to obtain the pure nebular spectra using the stellar population synthesis \textsc{starlight}
code \citep{2005MNRAS.358..363C, 2006MNRAS.370..721M, 2016MNRAS.460.1739V}. The emission lines were fitted using the publicly available \textsc{ifscube} package \citep{2020zndo...3945237R, 2021MNRAS.507...74R}. The fluxes were corrected for extinction following the procedure described by \citet{2021MNRAS.501.4064R},
where the observational H$\alpha$/H$\beta$ line ratio was compared with  
the theoretical value  (H$\alpha$/H$\beta$)=2.86 proposed by
\citet{1987MNRAS.224..801H} for a temperature of 10\,000 K and an electron density of 100 $\rm cm^{-3}$. From the resulting sample, we selected only the objects which present the 
[\ion{O}{ii}]$\lambda3726,\lambda3729$ (hereafter [\ion{O}{ii}]$\lambda3727$),
[\ion{O}{iii}]$\lambda4363$, 
H$\beta$, [\ion{O}{iii}]$\lambda5007$,  
H$\alpha$, [\ion{N}{ii}]$\lambda6584$, [\ion{S}{ii}]$\lambda6716,\lambda6731$, and 
[\ion{S}{iii}]$\lambda9069$ emission-lines with a signal to noise ratio (S/N) higher than 2.0.
Although the
presence of the [\ion{N}{ii}]$\lambda5755$ and  [\ion{S}{iii}]$\lambda6312$ auroral lines was not considered as selection criteria, when detected with (S/N) > 2 in the archival public data, their intensities were compiled.

Additionally, we also compiled from the  literature
emission-line intensities of Sy~2 nuclei  obtained by different authors and applying the same selection criteria used 
for the SDSS data, with exception of the presence of the [\ion{S}{iii}]$\lambda9069$ line, which is not measured in most of the available data. 
In these cases, a cross-correlation was performed between the objects with optical data and those whose [\ion{S}{iii}]$\lambda9069$ was measured by \citet{2006A&A...457...61R}, who presented a near-infrared (0.8-2.4 $\mu$m interval) spectral atlas of 47 AGNs. Initially, for each selected object, the
[\ion{S}{iii}]$\lambda9069$ fluxes from \citet{2006A&A...457...61R} 
were divided by the Pa$\beta$ flux. Thereafter,
in order to obtain the
[\ion{S}{iii}]$\lambda9069$ in relation to 
H$\beta$,  the (Pa$\beta$/H$\beta)=0.162$ theoretical ratio \citep{1989agna.book.....O} was assumed for a temperature of 10\,000 K and an electron density of 100 $\rm cm^{-3}$. A similar procedure was performed by \citet{2012A&A...547A..29B}.

Finally, for the entire sample, we applied  the criterion 
proposed by \citet{2001ApJ...556..121K} 
\begin{equation} 
\label{eq1}
\rm log([O\:III]\lambda5007/H\beta) \: > \: \frac{0.61}{log([N\:II]\lambda6584/H\alpha)-0.47}+1.19,
\end{equation}
to separate SF-like and AGN-like objects and the criterion proposed by \citet{2010MNRAS.403.1036C}
\begin{equation} 
\label{eq2}
\rm log([O\:III]\lambda5007/H\beta) \: > \: 0.47+log([N\:II]\lambda6584/H\alpha)\times1.10,
\end{equation}
to  separate AGN-like and  low-ionization nuclear emission-line region (LINER) objects.
The final sample resulted in 45 Sy~2 nuclei, which  is comprised by  33 objects  from SDSS data set
(redshift $z\: < \: 0.08$) and 12 objects from the literature
(redshift $z\: < \: 0.04$). 

The reduced number of objects
(33) resulting from  the SDSS database is mainly due to two selection criteria. First, the requirement for  [\ion{O}{iii}]$\lambda4363$ measured at $\rm (S/N)\: > \: 2$
makes it possible to select only 110 objects from a total of 333 Seyfert~2 nuclei.  A similar sample size  was obtained by \citet{2020MNRAS.496.2191F},
 where the  $T_{\rm e}$-method was applied in only 180 objects (see also \citealt{2012MNRAS.427.1266V, 2020MNRAS.492..468D})  selected from the SDSS DR8 \citep{2011ApJS..193...29A}.
Second, the requirement for the presence of both [\ion{O}{ii}]$\lambda3727$ and  [\ion{S}{iii}]$\lambda9069$  reduced our sample from 110 to only  33 objects.
 \citet{2006A&A...448..955I},  who  considered the  SDSS DR3 \citep{2005AJ....129.1755A}
database to estimate elemental abundances in SFs, also reported the difficulty in measuring
both [\ion{O}{ii}]$\lambda3727$  and [\ion{S}{iii}]$\lambda9069$ lines, mainly for 
 galaxies at $z \: \la \: 0.02$.  

In Figure~\ref{fig1},  an example of a pure Sy~2 nebular spectrum (in black) from the SDSS sample and the fitting (in red) produced by the \textsc{ifscube} package are shown. In Table~\ref{tab1} the reddening corrected emission line intensities
(in relation to H$\beta$=1.0) and the literature references from  which the data were compiled are listed. In this Table,  the theoretical relation $I(\lambda9069)=0.40 \times I(\lambda9532)$ between the [\ion{S}{iii}] emission-lines is assumed.

In Fig.~\ref{fig2}, a log([\ion{O}{iii}]$\lambda 5007$/H$\beta$) vs.\ log([\ion{N}{ii}]$\lambda 6584$/H$\beta$) diagnostic diagram, the observational data 
and the  above criteria (Equations~\ref{eq1} and \ref{eq2})  are shown.
It  can be seen that the Sy~2 sample
covers a large range of ionization degree
and metallicity, since a wide range of [\ion{O}{iii}]/H$\beta$ 
and [\ion{N}{ii}]/H$\alpha$ line ratio intensities are seen (e.g. \citealt{2006MNRAS.371.1559G, 2016MNRAS.456.3354F, 2020MNRAS.492.5675C}).

The observational data sample is heterogeneous, in the sense that the spectra were obtained with distinct instrumentation (e.g. long-slit, fiber spectroscopy), aperture, reddening correction procedures, etc.
These features could produce artificial  scattering or biases in the derived abundances.
\citet{2020MNRAS.492..468D, 2021MNRAS.501.1370D} and \citet{2021MNRAS.508..371A}  presented a complete discussion on the use of a heterogeneous sample and its possible implications on abundance estimates. These authors pointed out that the effects of considering 
such a heterogeneous sample on abundance
estimates produce uncertainties of $\sim0.1$ dex, i.e.
in the same order  or even lower than those derived by applying the $T_{\rm e}$-method
(e.g. \citealt{2003ApJ...591..801K, 2008MNRAS.383..209H}) and
strong-line methods (e.g. \citealt{1998AJ....115..909S, 2002MNRAS.330...69D}). Moreover, \citet{2005PASP..117..227K} presented a detailed analysis of the effect of considering different apertures on the determinations of physical parameters of galaxies. These authors
found that for
aperture capturing less than 20 per cent of the total galaxy emission,
the derived metallicity can differ by a factor of about 0.14 dex from the
value obtained when the total galaxy emission is considered.
However, only abundances of the nuclear regions are being considered here; therefore, the aperture effect on our  estimates is not  significant.
Additional analysis of uncertainties in abundance estimates derived from distinct instrumentation and/or aperture has been addressed, for instance,  by \citet{2021MNRAS.508.1582M}, who analysed
the Diffuse Ionized Gas (DIG) contribution to the nebular emission of SFs. These authors, specifically, found that the [\ion{S}{ii}] line fluxes tend to be more affected in comparison with other optical line fluxes.  \citet{2021MNRAS.508.1582M} also found that when spectra of local \ion{H}{ii} regions are extracted using large enough apertures while still avoiding the DIG, the observed line ratios are the same as in more distant galaxies. 
Therefore, there should not be any bias in our sample as a result of the usage of different instruments  (see also \citealt{2022ApJ...935...74A, 2022arXiv220913967P}). 
However, the requirement for the presence
of the weak  [\ion{O}{iii}]$\lambda4363$ line (about 100 times weaker than H$\beta$) in the SDSS spectra yields
a bias in our analysis, in the sense that objects with very high metallicity, where the gas suffers strong cooling and the electron temperature is low enough not to produce significant emission of this line, are mostly excluded. In fact, for instance, 
\citet{2020MNRAS.492..468D} selected from the SDSS DR7
database \citep{2009ApJS..182..543A}, 463 confirmed Sy2 spectra with only  150 objects  having
[\ion{O}{iii}]$\lambda4363$ measured with $\rm (S/N)\: > \: 2$ 
and from these, only 36/150 have oversolar metallicity according to the $T_{\rm e}$-method applied by \citet{2020MNRAS.496.3209D}.
Thus,  abundance determinations obtained in the present study do not extend to objects with the highest expected metallicity (see also  \citealt{1998AJ....116.2805V, 2006A&A...448..955I, 2020MNRAS.496.2191F}).

Another issue is the SF emission contribution to our AGN spectra, which can have a greater impact on the observed line fluxes   for the most distant objects.  \citet{2014MNRAS.439.3835D} presented a spatially resolved study of the active galaxy NGC\,7130 ($z=0.016$) and found
that SFs are responsible for 30 and 65 per cent
of the [\ion{O}{iii}] and H$\alpha$ luminosity, respectively.
Moreover, \citet{2022arXiv221113648V}
compared results from the  NLR photoionization models
\citep{2016MNRAS.456.3354F} incorporated into the \textsc{Beagle} (Bayesian SED-fitting, \citealt{2016MNRAS.462.1415C}) code with observational  spectroscopic data and showed that the SF H$\beta$  flux contribution  to the total nuclear flux of an active galaxy can range from 0 to 50  per cent. However, we emphasize that, in principle, the SF flux contribution has a minimal effect on AGN abundance estimates when a sample of objects is considered. This assertion is supported by \citet{2019ApJ...874..100T}, who demonstrated that the aperture effect (and consequently SF contribution) has a negligible impact on metallicity estimates once  comparable mass-metallicity relations for galaxies in  four redshift bins were considered.

\begin{figure}
\includegraphics[angle=0.0,width=0.47\textwidth]{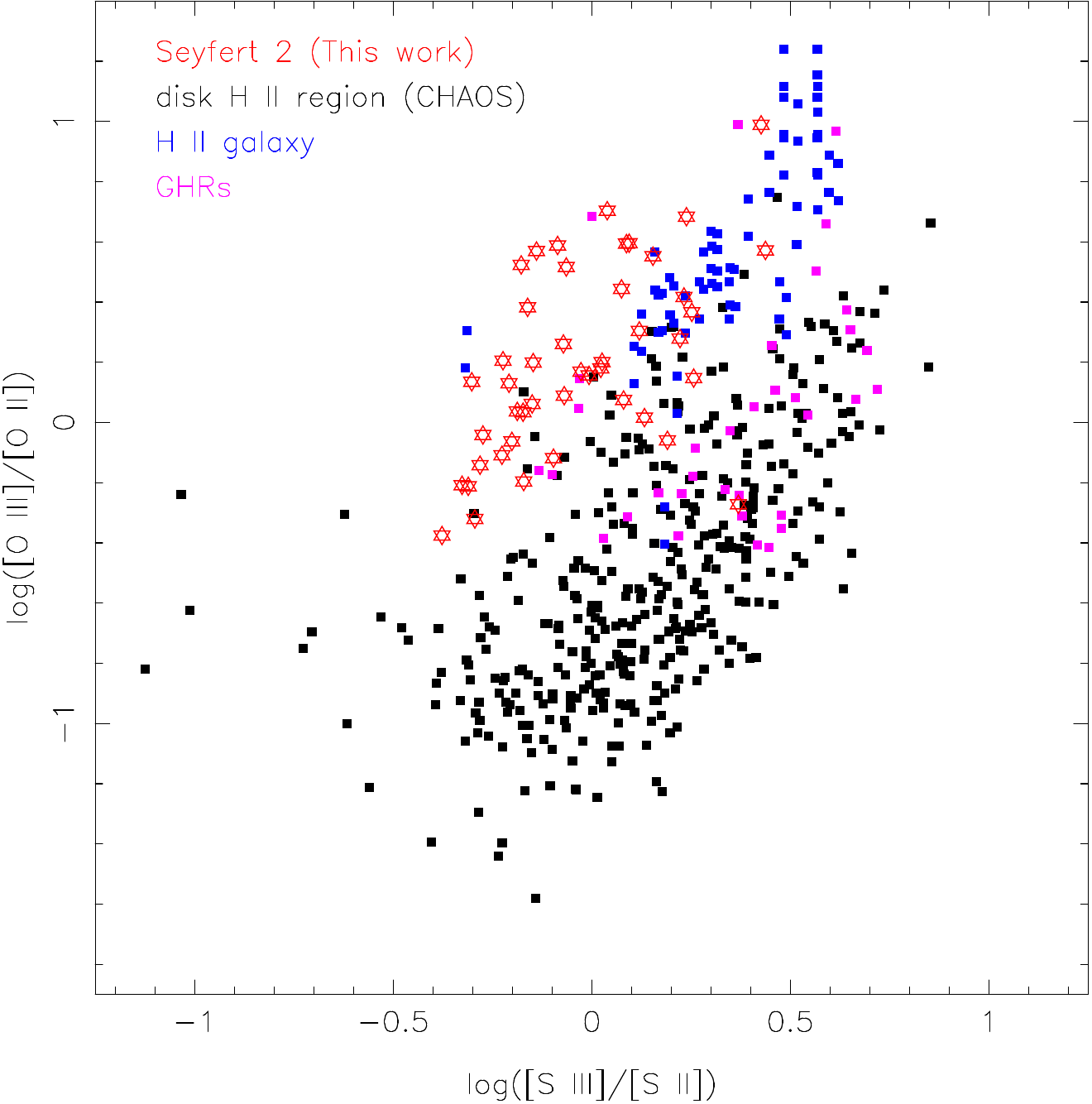}
\caption{Logarithm of [\ion{O}{iii}]$\lambda5007$/[\ion{O}{ii}]$\lambda3727$ versus
[\ion{S}{iii}]$\lambda9069+\lambda9532$/[\ion{S}{ii}]$\lambda6716+\lambda6731$. Red points represent the Sy~2 sample (see Sect.~\ref{observ}) whose emission line intensities are listed in Table~\ref{tab1}. Black points represent
disk \ion{H}{ii} regions whose data were taken from the \textsc{chaos} project
and obtained by \citet{2015ApJ...806...16B, 2020ApJ...893...96B}, \citet{2015ApJ...808...42C, 2016ApJ...830....4C} and \citet{2021ApJ...915...21R, 2022arXiv220903962R}.
Blue points represent \ion{H}{ii} galaxies taken from \citet{2006MNRAS.372..293H,2008MNRAS.383..209H,2011MNRAS.414..272H,2012MNRAS.422.3475H} and pink points represent GHRs taken from \citet{2006MNRAS.372..293H}.  
}
\label{fig3}
\end{figure}

Since the [\ion{S}{iii}] lines of Seyfert galaxies are rarely found in the literature due to the fact that they are located in the near infrared which there are few instruments operating, it is worthwhile to compare their emission line flux ratios with those  of SFs.
In this regard,  we consider  emission line intensities of 
\ion{H}{ii} galaxies (44 objects) and Giant \ion{H}{ii} regions (GHRs, 34 objects) presented
by \citet{2006MNRAS.372..293H,2008MNRAS.383..209H,2011MNRAS.414..272H,2012MNRAS.422.3475H}.  
Besides, we compare the Sy~2 emission lines considered in this work with those from 378 disk \ion{H}{ii} regions located in 6 local spiral galaxies, which have been made available by the \textsc{CHAOS} project\footnote{\url{https://www.danielleaberg.com/chaos}}, 
and presented by \citet{2015ApJ...806...16B, 2020ApJ...893...96B}, \citet{2015ApJ...808...42C, 2016ApJ...830....4C}, and \citet{2021ApJ...915...21R, 2022arXiv220903962R}. This comparison (see also \citealt{2022MNRAS.511.4377D}) is shown in Fig.~\ref{fig3}.  A clear correlation is derived between the two dataset since both line ratios are dependent on the ionization degree of the gas. Interestingly, Sy~2 nuclei present similar [\ion{S}{iii}]/[\ion{S}{ii}] line ratio intensities to those of SFs.
The Sy~2 [\ion{O}{iii}]/[\ion{O}{ii}] line ratio intensities are in consonance with those of \ion{H}{ii} galaxies and GHRs  and are higher than those from disk \ion{H}{ii} regions.

\section{Abundance estimates}
\label{abund}

\begin{table*} 
\centering 
\caption{Atomic dataset used for  collisionally excited lines of selected  element ions.} 
\label{tatomic} 
\begin{tabular}{@{}lcc@{}} 
\hline 
Ion                       &             Transition probabilities                     &  Collisional strengths       \\ 
\hline 
S$^{+}$  & \citet{2006ADNDT..92..607F} & \citet{2010ApJS..188...32T} \\ 
S$^{2+}$ & \citet{2006ADNDT..92..607F} &   \citet{1999ApJ...526..544T}  \\
O$^{+}$  & \citet{1996atpc.book.....W} & \citet{2009MNRAS.397..903K}  \\ 
O$^{2+}$ & \citet{2004ADNDT..87....1F}, \citet{2000MNRAS.312..813S} & \citet{2014MNRAS.441.3028S} \\
\hline 
\end{tabular} 
\end{table*}

\begin{table*}
\caption{Estimates of $T_{\rm e}(\rm O^{2+})$, $T_{\rm e}(\rm N^{+})$ 
and $T_{\rm e}(\rm S^{2+})$
used to obtain the relations given by Eqs.~\ref{eq3} and \ref{eq4}. 
The line ratios are: $RO3=\azul{(}[\ion{O}{iii}]\lambda4959+\lambda5007)/\lambda4363$, $RN2=\azul{(}[\ion{N}{ii}]\lambda6548+\lambda6584)/\lambda5755$,
$RS3=\azul{(}[\ion{S}{iii}]\lambda9069+\lambda9532)/\lambda6312$ and
$RS2=[\ion{S}{ii}]\lambda6716/\lambda6731$.}
\label{tab0}
\begin{tabular}{lccccccccc}	 
\noalign{\smallskip} 
\hline 
 Object        &	$RO3$  	    &	   $RN2$  	 &     $RS3$        &    $RS2$		 & $T_{\rm e}$($\rm O^{2+}$)[K] & $T_{\rm e}$($\rm N^{+}$)[K] & $T_{\rm e}$($\rm S^{2+}$)[K]	  & $N_{\rm e}[\rm cm^{-3}]$ &   Ref. \\ 
\hline
IZw\,92        & 42.62 $\pm$ 8.52   & 66.50 $\pm$ 13.30  &  ---             & 0.92 $\pm$ 0.09 	 & 16350 $\pm$ 1646                    & 9929 $\pm$ 788                      &  ---  					  & 1176 $\pm$ 427   & 1 \\
Mrk\,3         & 69.41 $\pm$ 13.88  & 87.29 $\pm$ 17.45  &  ---             & 0.89 $\pm$ 0.08 	 & 13151 $\pm$ 1059                    & 8969 $\pm$ 638 		     &  ---  					  & 1221 $\pm$ 395   & 2 \\
Mrk\,78        & 112.90 $\pm$ 22.58 & 95.31 $\pm$ 19.06  &  ---             & 1.11 $\pm$ 0.11 	 & 11023 $\pm$ 743                     & 8754 $\pm$ 607 		     &  ---  					  & 467 $\pm$ 229    & 2 \\
Mrk\,34        & 100.90 $\pm$ 20.18 & 87.27 $\pm$ 17.45  &  ---             & 1.02 $\pm$ 0.10 	 & 11445 $\pm$ 800                     & 9012 $\pm$ 643 		     &  ---  					  & 691 $\pm$ 280    & 2 \\
Mrk\,1         & 69.09 $\pm$ 13.81  & 135.45 $\pm$ 27.09 &  ---             & 0.94 $\pm$ 0.09 	 & 13187 $\pm$ 1060                    & 7773 $\pm$ 481 		     &  ---  					  & 1000 $\pm$ 362   & 2 \\
Mrk\,533       & 95.00 $\pm$ 19.00  & 240.00 $\pm$ 48.00 &  ---             & 0.86 $\pm$ 0.08 	 & 11681 $\pm$ 834                     & 6591 $\pm$ 346 		     &  ---  					  & 1319 $\pm$ 429   & 3 \\
Mrk\,612       & 60.00 $\pm$ 12.00  & 112.00 $\pm$ 22.40 &  ---             & 1.36 $\pm$ 0.13 	 & 13993 $\pm$ 1199                    & 8304 $\pm$ 551 		     &  ---  					  & 87 :             & 3 \\
ESO\,138\,G1   & 34.23 $\pm$ 6.84   & 47.73 $\pm$ 9.54   &  ---             & 0.97 $\pm$ 0.09 	 & 18386 $\pm$ 2084                    & 11457 $\pm$ 1055                    &  ---  					  & 1003 $\pm$ 366   & 4 \\
NGC 2992       & 40.36 $\pm$ 8.07   & 98.20 $\pm$ 19.64  &  ---             & 1.12 $\pm$ 0.11 	 & 16831 $\pm$ 1736                    & 8659 $\pm$ 595 		     &  ---  					  & 514 $\pm$ 266    & 5 \\
NGC 2210       & 24.36 $\pm$ 4.87   & 117.77 $\pm$ 23.55 &  ---             & 1.06 $\pm$ 0.10 	 & 22797 $\pm$ 3095                    & 8143 $\pm$ 527 		     &  ---  					  & 744 $\pm$ 328    & 5 \\
NGC\,5506      & 73.57 $\pm$ 14.70  & 168.50 $\pm$ 33.70 &  ---             & 0.92 $\pm$ 0.09 	 & 12859 $\pm$ 1012                    & 7279 $\pm$ 421 		     &  ---  					  & 1074 $\pm$ 386   & 5 \\
Mrk\,348       & 23.7  $\pm$ 2.4    & 31.8   $\pm$ 11.9  &                  & 0.83 $\pm$ 0.03    & 28600 $\pm$ 4700                    & 21700 $\pm$ 6100                    &  ---                                       & 1940 $\pm$ 245  & 6 \\
Mrk\,607       & 29.7  $\pm$ 10.2   & 80.6   $\pm$ 16.2  &                  & 0.87 $\pm$ 0.11    & 23500 $\pm$ 2700                    & 10200 $\pm$ 1300                    &  ---                                       & 1548 $\pm$ 707  & 6 \\
56067-0382     & 41.43 $\pm$ 8.29   &  12.29 $\pm$ 2.46  &  ---             & 1.06 $\pm$ 0.16 	 & 16604 $\pm$ 1694                    & 31367 $\pm$ 8555 		     &  ---  					  & 669 $\pm$ 463    & 7 \\
55539-0167     & 82.20 $\pm$ 16.44  & 24.49 $\pm$ 4.90   &  ---             & 1.13 $\pm$ 0.17 	 & 12334 $\pm$ 928                     & 16655 $\pm$ 2244 		     &  ---  					  & 442 $\pm$ 349    & 7 \\
56001-0293     & 109.00 $\pm$ 21.80 & 37.02 $\pm$ 7.40   &  ---             & 1.39 $\pm$ 0.21 	 & 11159 $\pm$ 760                     & 13147 $\pm$ 1393 		     &  ---  					  & 58 :             & 7 \\
55742-0383     & 99.80 $\pm$ 19.96  & 42.04 $\pm$ 8.41   &  ---             & 1.13 $\pm$ 0.17 	 & 11497 $\pm$ 806                     & 12245 $\pm$ 1202 		     &  ---  					  & 432 $\pm$ 340    & 7 \\
55302-0655     & 41.20 $\pm$ 8.24   & 47.71 $\pm$ 9.54   &  ---             & 1.24 $\pm$ 0.19 	 & 16672 $\pm$ 1708                    & 11537 $\pm$ 1068 		     &  ---  					  & 266 $\pm$ 0      & 7 \\
56568-0076     & 105.33 $\pm$ 21.07 & 96.13 $\pm$ 19.23  &  ---             & 1.03 $\pm$ 0.15 	 & 11280 $\pm$ 780                     & 8715 $\pm$ 604                      &  ---  					  & 660 $\pm$ 407    & 7 \\
56566-0794     & 94.80 $\pm$ 18.96  & 127.06 $\pm$ 25.41 & 4.36 $\pm$ 0.87  & 1.18 $\pm$ 0.18 	 & 11713 $\pm$ 838                     & 7968 $\pm$ 505                      & 21665 $\pm$ 3991                           & 342 :            & 7 \\
55181-0154     & 130.33 $\pm$ 26.07 &  ---               & 3.85 $\pm$ 0.77  & 0.98 $\pm$ 0.15 	 & 10508 $\pm$ 675                     &  ---                                & 24380 $\pm$ 5015                           & 786 $\pm$ 474    & 7 \\
56088-0473     & 128.33 $\pm$ 25.67 &  ---               & 18.00 $\pm$ 3.60 & 1.09 $\pm$ 0.16 	 & 10566 $\pm$ 682                     &  ---   			     & 9244 $\pm$ 747	                          & 502 $\pm$ 349    & 7 \\
56034-0154     & 84.86 $\pm$ 16.97  &  ---               & 12.00 $\pm$ 2.40 & 1.09 $\pm$ 0.16 	 & 12184 $\pm$ 909                     &  ---   			     & 11052 $\pm$ 1070                           & 527 $\pm$ 370    & 7 \\
56626-0636     & 102.00 $\pm$ 20.40 &  ---               & 3.49 $\pm$ 0.70  & 1.08 $\pm$ 0.16 	 & 11409 $\pm$ 795                     &  ---   			     & 27079 $\pm$ 6116                           & 537 $\pm$ 371    & 7 \\
55651-0052     & 177.00 $\pm$ 35.40 &  ---               & 5.37 $\pm$ 1.07  & 1.06 $\pm$ 0.16 	 & 9576 $\pm$ 560                      &  ---   			     & 18131 $\pm$ 2884                           & 549 $\pm$ 368    & 7 \\
56206-0454     & 74.11 $\pm$ 14.82  &  ---               & 6.17 $\pm$ 1.23  & 1.23 $\pm$ 0.18 	 & 12842 $\pm$ 1010                    &  ---   			     & 16315 $\pm$ 2354                           & 267 :            & 7 \\
55860-0112     & 71.09 $\pm$ 14.22  &  ---               & 3.39 $\pm$ 0.68  & 1.02 $\pm$ 0.15 	 & 13038 $\pm$ 1038                    &  ---   			     & 27975 $\pm$ 6506                           & 724 $\pm$ 448    & 7 \\
55710-0116     & 107.71 $\pm$ 21.54 &  ---               & 12.29 $\pm$ 2.46 & 0.92 $\pm$ 0.14 	 & 11189 $\pm$ 764                     &  ---   			     & 10917 $\pm$ 1047                           & 1018 $\pm$ 578   & 7 \\
56366-0928     & 47.50 $\pm$ 9.50   &  ---               & 3.14 $\pm$ 0.63  & 1.00 $\pm$ 0.15 	 & 15523 $\pm$ 1474                    &  ---   			     & 30648 $\pm$ 7717                           & 834 $\pm$ 517    & 7 \\
56328-0550     & 81.67 $\pm$ 16.33  &  ---               & 7.48 $\pm$ 1.50  & 1.28 $\pm$ 0.19 	 & 12373 $\pm$ 934                     &  ---   			     & 14339 $\pm$ 1821                           & 190 :            & 7 \\
55617-0758     & 76.80 $\pm$ 15.36  &  ---               & 3.45 $\pm$ 0.69  & 0.90 $\pm$ 0.13 	 & 12640 $\pm$ 978                     &  ---   			     & 27394 $\pm$ 6222 			  & 1156 $\pm$ 609   & 7 \\
56003-0218     & 67.14 $\pm$ 13.43  &  ---               & 5.89 $\pm$ 1.18  & 0.97 $\pm$ 0.14 	 & 13337 $\pm$ 1088                    &  ---   			     & 16867 $\pm$ 2531 			  & 889 $\pm$ 504    & 7 \\
55505-0654     & 198.75 $\pm$ 39.75 &  ---               & 12.00 $\pm$ 2.40 & 1.12 $\pm$ 0.17 	 & 9268 $\pm$ 524                      &  ---   			     & 11052 $\pm$ 1072 			  & 422 $\pm$ 319    & 7 \\
\hline
\end{tabular}	   									
\begin{minipage}[c]{2\columnwidth}
References: (1) \citet{1994ApJ...435..171K}, (2) \citet{1978ApJ...223...56K}, (3) \citet{1981ApJ...250...55S}, 
(4) \citet{1992A&A...266..117A}, (5) \citet{1980ApJ...240...32S}, (6), \citet{2021MNRAS.501L..54R}, (7) SDSS sample.
\end{minipage}
\end{table*} 

For the Sy~2 sample previously described, we determined the sulphur  and oxygen abundances relative to hydrogen.  To do that,  
electron temperatures representing  the zones where distinct ions are located in the gas phase,
electron density and ionic abundances were calculated
using the  1.1.13 version of {\sc PyNeb} code \citep{Luridiana2015}, which permits an interactive procedure in the derivation of these parameters. The references for the predefined atomic parameters incorporated into the {\sc PyNeb} code are listed in Table~\ref{tatomic}.

As the line measurements for some objects (9/45) of our sample does not present observational errors, the abundance uncertainties were estimated using Monte Carlo simulations.
For each diagnostic line, we generate 1000 random values assuming a Gaussian distribution with a standard deviation equal to the associated uncertainty of 10\,\% and 20\,\% for strong (e.g. [\ion{O}{iii}]$\lambda5007$) and auroral (e.g. [\ion{O}{iii}]$\lambda4363$) line intensities involved in the diagnostics, respectively.
Thereafter, an empirical Ionization Correction Factor (ICF) was considered in the derivation of the total sulphur  abundance. 
For objects  that have measured emission-line errors (36/45), the uncertainties in the
final abundance values were obtained propagating the  errors in the line measurements, electron temperature and electron density.
Subsequently, the description
of the employed methodology is presented.

\subsection{Temperature estimations \label{sectene}}
 
Several  studies have been directed to  estimate the chemical composition of 
SFs and, in almost all of these estimates, it has been a common practice to use temperature relations derived from photoionization models to infer the temperatures in the unobserved ionization zones (e.g. \citealt{1990A&AS...83..501S, 1992AJ....103.1330G, 2003MNRAS.346..105P, 2006A&A...448..955I}). However, when temperature relations predicted
by  photoionization models simulating SFs are compared with direct estimates relying on auroral lines, large deviations are found, reaching up to $\sim 5000$ K (e.g. \citealt{2008MNRAS.383..209H,2020ApJ...893...96B, 2020MNRAS.497..672A}). 
Despite the fact that temperature relations for AGNs are barely found in the literature (see 
\citealt{2020MNRAS.496.3209D, 2021MNRAS.508..371A, 2021MNRAS.508.3023M}), it seems that
similar disagreement is also derived for this class of object. In fact, \citet{2021MNRAS.501L..54R} 
compared the $T_{\rm e}(\rm N^{+})$-$T_{\rm e}(\rm O^{2+})$
relation predicted by
photoionization models, built using 
the \textsc{cloudy} code \citep{2013RMxAA..49..137F}, 
with values derived from observational auroral emission lines for a sample of 12 local Seyfert nuclei.
The model predictions reproduce the direct temperature observations for all objects, except for Mrk\,348 for which the direct $T_{\rm e}(\rm N^{+})$  value is $\sim$10\,000 K higher than the predicted one. This object is know to host ionized gas outflows \citep{2018MNRAS.476.2760F} and probably the higher observed temperatures are due to extra heating caused by shocks (see \citealt{2021MNRAS.501.1370D}), which is not  accounted  in the photoionization models considered 
by \citet{2021MNRAS.501L..54R}.  
Obviously, additional
comparison with a larger sample of objects combined with kinematic studies (e.g. \citealt{2021MNRAS.506.2725X}; Flury et al. 2022, in preparation) of objects where gas
outflow is detected, is necessary to confirm this result.  

 Since  comparisons between  
 observational emission line intensities and the ones predicted by photoionization models   
  indicate that the main ionization source of  most NLR of Sy~2s
 is the radiation emitted by gas accretion into a supermassive black hole, for objects in the
 local universe (see e.g.\ \citealt{1984A&A...135..341S,1986ApJ...300..658F,1998AJ....115..909S,2006MNRAS.371.1559G,2016MNRAS.456.3354F,2017MNRAS.467.1507C,2017MNRAS.468L.113D,2020MNRAS.496.3209D,2019MNRAS.489.2652P,2019ApJ...874..100T,2020MNRAS.492.5675C,2021MNRAS.508..371A}) and also at high redshift (see e.g.\ \citealt{2006A&A...459...85N, 2009A&A...503..721M, 2018A&A...616L...4M,2014MNRAS.443.1291D,2018MNRAS.479.2294D,2019MNRAS.486.5853D,2018A&A...612A..94N,2019A&A...626A...9M,2020ApJ...898...26G}), we are able to apply the $T_{\rm e}$-method to derive reliable estimates. However, the weak temperature-sensitive auroral emission-line measurements are barely available in the literature for AGNs, therefore we developed our own empirical method based on our sample.

\begin{figure}
\includegraphics[angle=0.0,width=0.47\textwidth]{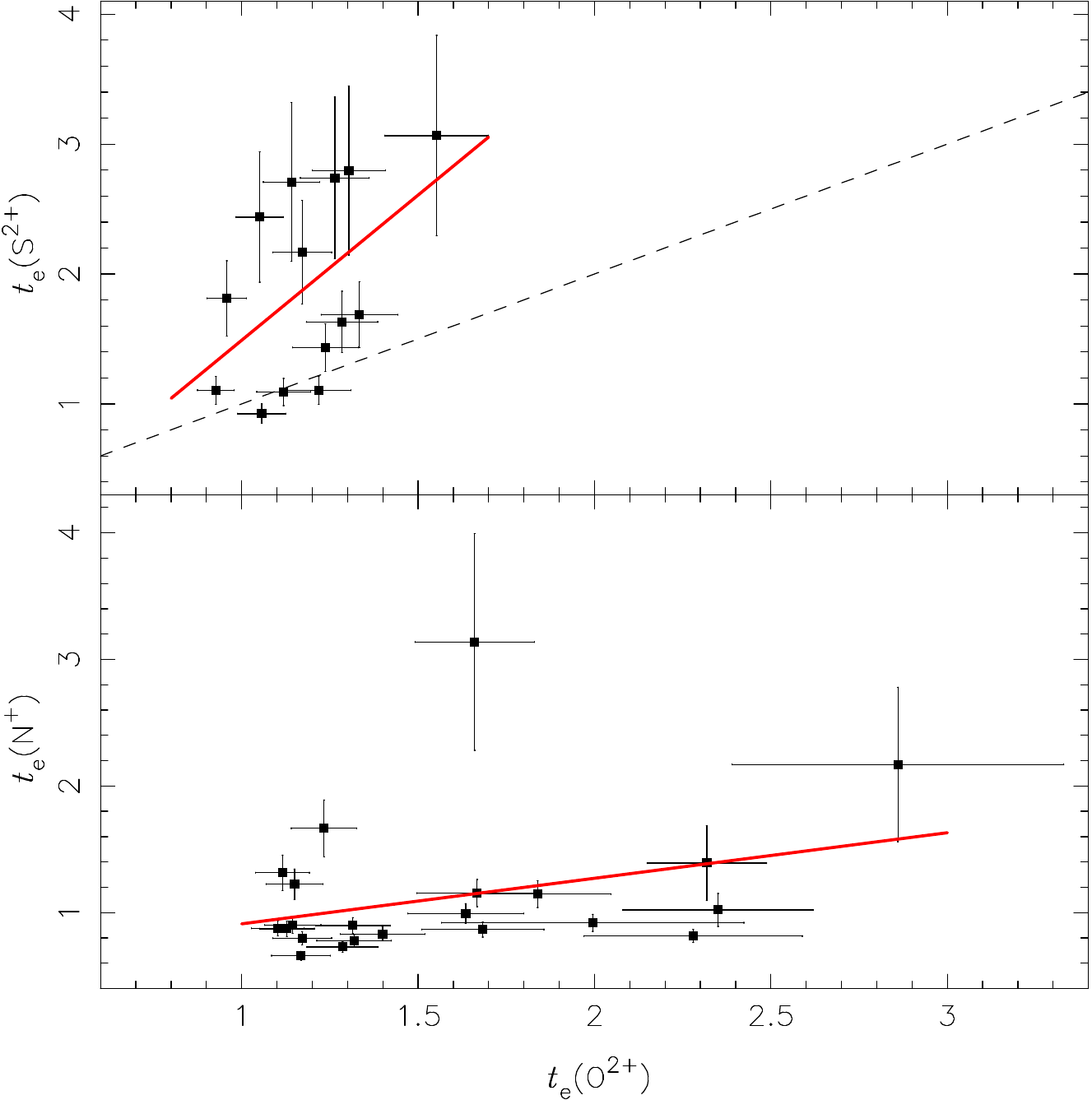}
\caption{Comparing direct temperature values for different ions. Values are in units of $10^{4}$ K. Botton panel: $t_{\rm e}(\rm O^{2+})$ and $t_{\rm e}(\rm N^{+})$  are derived through the $RO3$=[\ion{O}{iii}]($\lambda4959+\lambda5007$)/$\lambda4363$ and $RN2$=[\ion{N}{ii}]($\lambda6548+\lambda6584$)/$\lambda5755$ line intensities ratios, respectively, and the electron density (from the $RS2$=[\ion{S}{ii}]$\lambda6716/\lambda6731$) listed in Table~\ref{tab0}, and
by using version 1.1.13 of {\sc PyNeb} code \citep{Luridiana2015}. 
The  red line represents the linear regression to the points given by Eq.~\ref{eq3}.
Top panel: same as bottom panel but  for
$t_{\rm e}(\rm S^{2+})$ values  derived from $RS3$=[\ion{S}{iii}]($\lambda9069+\lambda9532$)/$\lambda6312$.
Red line represents the linear regression to the points given by Eq.~\ref{eq4} while the dashed line the equality between the estimates.}
\label{fig4}
\end{figure}

To derive the empirical relations for our abundance estimates, we used  auroral  line intensities from our sample  and those of Sy~2 NLRs  available in the literature. Firstly, the observational intensities of
the $RO3$=[\ion{O}{iii}]($\lambda4959+\lambda5007$)/$\lambda4363$ and 
$RN2$=[\ion{N}{ii}]($\lambda6548+\lambda6584$)/$\lambda5755$ line ratios were used to
calculate $T_{\rm e}(\rm O^{2+})$ and $T_{\rm e}(\rm N^{+})$, respectively, for 18 objects, i.e. 7 objects of our sample (see Table~\ref{tab1})  and 11  Sy~2 compiled by \citet{2020MNRAS.496.3209D}.
To derive  $T_{\rm e}(\rm S^{2+})$, we used the
$RS3$=[\ion{S}{iii}]($\lambda9069+\lambda9532$)/$\lambda6312$
line ratios for 14 objects (over 45) in our sample (see Table~\ref{tab1}). For each object, the 
temperature estimates were performed assuming a constant electron density ($N_{\rm e}$) value across the nebula, which is derived from the $RS2$=[\ion{S}{ii}]$\lambda6716/\lambda6731$ intensity ratio.  In Table~\ref{tab0},
the objects and their corresponding  $RO3$, $RN2$, $RS3$ and $RS2$ line intensities ratios, the electron density, $T_{\rm e}(\rm O^{2+})$, $T_{\rm e}(\rm N^{+})$ and $T_{\rm e}(\rm S^{2+})$ temperature derived values  are listed. We note that the object 56067-0382 has a $T_{\rm e}(\rm N^{+})$
 higher  than those derived for other objects
and similar to the value derived for Mrk\,348 by \citet{2021MNRAS.501L..54R}.
Probably 56067-0382  presents  gas outflows but its temperatures were still considered. 
Since in most cases only the [\ion{O}{iii}]$\lambda4363$ auroral line is measured (e.g. \citealt{1998AJ....116.2805V, 2003ApJ...591..801K}), we proposed, as usual, temperature relations  with respect to $T_{\rm e}(\rm O^{2+})$. In Fig.~\ref{fig4},  $T_{\rm e}(\rm N^{+})$ and $T_{\rm e}(\rm S^{2+})$ are plotted against $T_{\rm e}(\rm O^{2+})$, with the values in units of $10^{4}$ K. In the upper panel of this figure, the dashed line represents the equality between 
$T_{\rm e}(\rm S^{2+})$ and $T_{\rm e}(\rm O^{2+})$.
As for SFs (e.g., see \citealt{2008MNRAS.383..209H, 2020ApJ...893...96B}), clear correlations between the Sy~2 temperatures are observed, with a linear regression resulting in 
 \begin{equation}
\label{eq3}
t_{\mathrm{e}}(\mathrm{N^{+}})=  0.36(\pm0.23)\times  t_{\mathrm{e}}(\mathrm{O^{2+}}) + {\rm 0.55(\pm0.39)},
\end{equation}
and

\begin{equation}
\label{eq4}
t_{\mathrm{e}}(\mathrm{S^{2+}})=  2.23(\pm1.12)\times t_{\mathrm{e}}(\mathrm{O^{2+}}) - {\rm 0.74(\pm1.34)},
\end{equation} 
where $t_{\rm e}=T_{\rm e}/10^{4} \: \rm K$.  

It can be seen in Fig.~\ref{fig4} that higher temperature values for
$\rm S^{2+}$ are derived in comparison with those for $\rm O^{2+}$, probably indicating that the former ion occupy an inner gas region than the latter. 
Conversely, an opposite result is derived for disk \ion{H}{ii} regions, 
i.e. $T_{\rm e}(\rm O^{2+})$ is $\sim1000$ K higher than $T_{\rm e}(\rm S^{2+})$ (e.g. \citealt{2021ApJ...915...21R}). \cite{2006MNRAS.372..293H} analysed the relation between $T_{\rm e}(\rm O^{2+})$ and $T_{\rm e}(\rm S^{2+})$ using a sample that comprises \ion{H}{ii} galaxies, giant extragalactic \ion{H}{ii} regions, Galactic \ion{H}{ii} regions, and \ion{H}{ii} regions from the Magellanic Clouds (MCs). These authors found that the [\ion{S}{iii}] electron temperatures are higher than the corresponding [\ion{O}{iii}] estimations for most objects presenting temperatures higher than about 14\,000K, mainly the metal poor \ion{H}{ii} galaxies, and the opposite behaviour for the coolest nebulae, mainly giant extragalactic \ion{H}{ii} regions, Galactic \ion{H}{ii} regions, and \ion{H}{ii} regions from the MCs, which present the highest metallicities. Taking into account the temperatures of the different samples studied by \citet[][\ion{H}{ii} galaxies that present the higher temperatures]{2008MNRAS.383..209H} and \citet[][galactic disk \ion{H}{ii} regions with lower temperatures]{2021ApJ...915...21R}, the $T_{\rm e}(\rm S^{2+})$-$T_{\rm e}(\rm O^{2+})$ behavior derived in the present paper for Sy~2s is the same as found by \cite{2006MNRAS.372..293H}.

\begin{table}
\caption{Critical densities for collisional de-excitation for the lines involved in the present work. Values were
calculated with the \textsc{PyNeb} code \citep{Luridiana2015} assuming an electron
temperature of 15\,000 K.}
\label{tabf}
\centering
\begin{tabular}{@{}lc@{}}
\hline
\noalign{\smallskip}
Line                         & $N_{\rm c}$ ($\rm cm^{-3})$ \\ 
\noalign{\smallskip}
$[\ion{O}{ii}]\lambda3726$   &  $4.72\times 10^{3}$      \\
$[\ion{O}{ii}]\lambda3729$   &  $1.49\times 10^{3}$      \\
$[\ion{O}{iii}]\lambda4363$  &  $2.88\times 10^{7}$      \\
$[\ion{O}{iii}]\lambda5007$  &  $7.83\times 10^{5}$      \\
$[\ion{N}{ii}]\lambda5755$   &  $1.87\times 10^{7}$      \\
$[\ion{S}{iii}]\lambda6312$  &  $1.44\times 10^{7}$      \\
$[\ion{N}{ii}]\lambda6484$   &  $1.04\times 10^{5}$      \\
$[\ion{S}{ii}]\lambda6716$   &  $1.37\times 10^{3}$      \\
$[\ion{S}{ii}]\lambda6731$   &  $3.67\times 10^{3}$      \\
$[\ion{S}{iii}]\lambda9069$  &  $6.42\times 10^{5}$      \\
\hline  	   											
\end{tabular}	   									
\end{table}

\begin{figure}
\includegraphics[angle=0.0,width=0.47\textwidth]{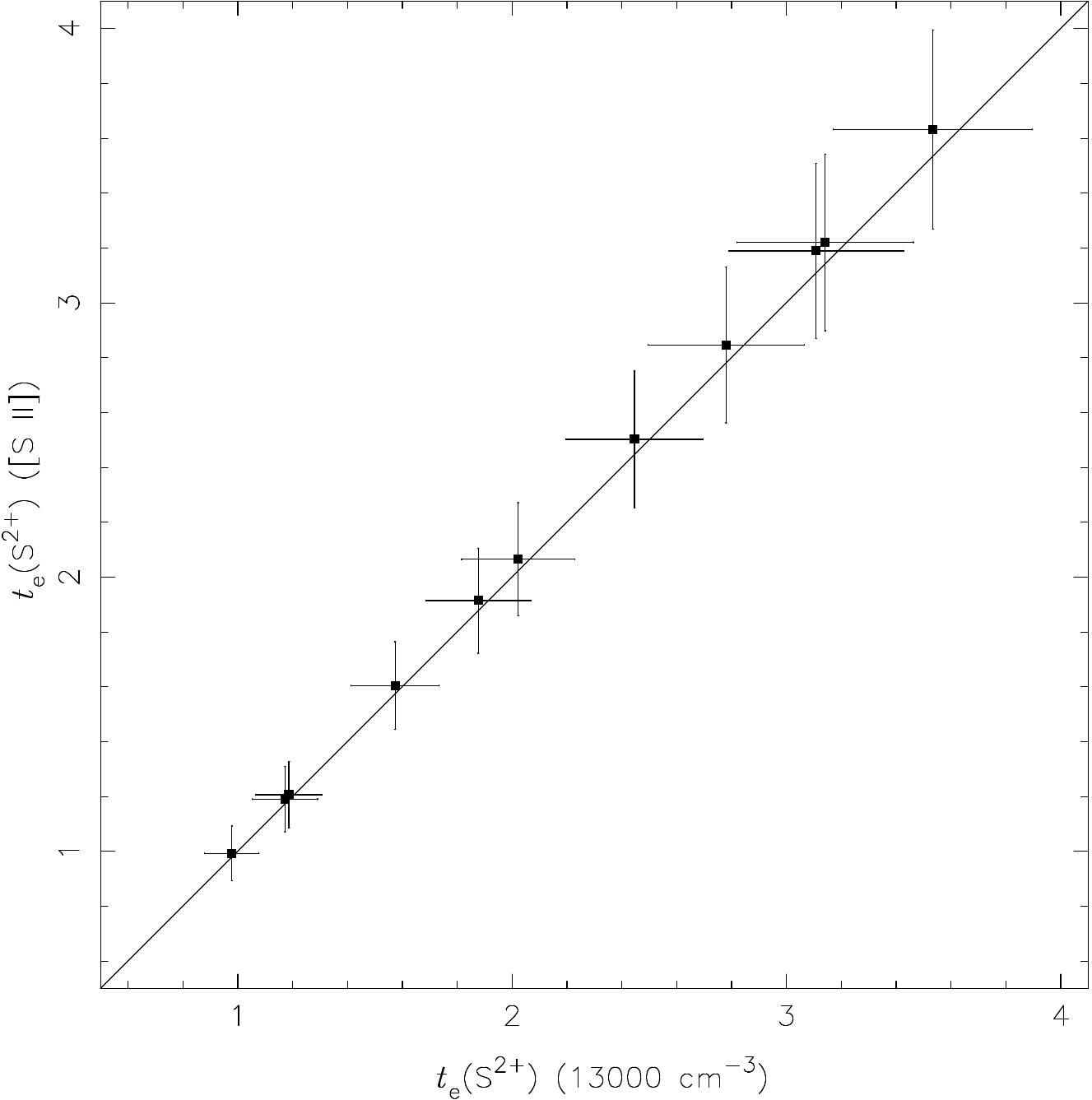}
\caption{Comparison between electron temperatures (in units of $10^{4}$ K) derived from the $RS3$=[\ion{S}{iii}]($\lambda9069+\lambda9532$)/$\lambda6312$ assuming electron density estimations via the $RS2$=[\ion{S}{ii}]$\lambda6716/\lambda6731$ (listed in Table~\ref{tab0}) with the estimates assuming a value of
$13\,000\, {\rm cm^{-3}}$ as derived by \citet{2017MNRAS.471..562C}
from [\ion{Ar}{iv}]$\lambda4711/\lambda4740$. Line represents the equality between the two estimates.}
\label{fig5}
\end{figure}

Spatially resolved observational studies of NLRs
have found a profile of electron density along the AGN radius, in the sense that denser gas is located in the inner regions.
For instance, \citet{2018MNRAS.476.2760F}, who obtained emission-line flux of two-dimensional maps from five bright nearby Seyfert nuclei,  obtained
electron densities ranging from $\sim 2500\: \rm cm^{-3}$ in the central
parts to $\sim 100\: \rm cm^{-3}$ in the outskirts (see also \citealt{2018ApJ...856...46R, 2018ApJ...867...88R, 2021ApJ...910..139R, 2022ApJ...930...14R, 2018A&A...618A...6K, 2019A&A...622A.146M, 2021MNRAS.507...74R}). Moreover, electron density estimations through the  
[\ion{Ar}{iv}]$\lambda4711/\lambda4740$ line ratio, which
traces the density in the innermost layers,    showed  values of up to
$\sim$13\,000 $\rm cm^{-3}$ (e.g. \citealt{2017MNRAS.471..562C, 2021MNRAS.500.2666C}). Thus, density values derived from [\ion{S}{ii}] lines may not be representative of the region where $\rm S^{2+}$ ions are located,  which could inherently introduce an error in the $T_{\rm e}(\rm S^{2+})$ values. In order to explore the influence of the electron density on the sulphur  temperature  estimations, we show in Fig.~\ref{fig5}  the $t_{\rm e}(\rm S^{2+})$ derived assuming the $N_{\rm e}$ values from [\ion{S}{ii}]$\lambda6716/\lambda6731$ line ratios (listed in Table~\ref{tab0}) versus the estimations considering a fixed value of 13\,000 $\rm cm^{-3}$, as derived by \citet{2017MNRAS.471..562C} for the extended narrow-line region of the Seyfert
2 galaxy IC\,5063. We notice a good agreement between the values, with a difference of $\sim2\,\%$, which is lower than the uncertainty produced by the error in the line measurements
($\sim 10\%$, see Table~\ref{tab0}). In Table~\ref{tabf} the
critical density ($N_{\rm c}$) values for the lines involved in the present work,
calculated with the \textsc{PyNeb} code \citep{Luridiana2015} assuming an electron
temperature of 15\,000 K, are listed. One can see that $N_{\rm c}$
values are higher than the electron density values (listed in Table~\ref{tab2}) derived for our Sy~2 sample.
Thus, electron density variations in NLRs have a minimal influence on  our temperature estimates.
For some objects, the emission-line errors are significant, thus frequently with density error
bars larger than the determinations themselves. Therefore we use ``:'' to indicate that error
bars are at least an order of magnitude larger than the expected density (see Table~\ref{tab2}).

Due to the similarity among the $\rm S^{+}$, 
$\rm N^{+}$ and $\rm O^{+}$ ionization potentials (23.33 eV, 29.60 eV and 35.12 eV, respectively) and because the  [\ion{S}{ii}]$\lambda\lambda4068,4076$ and [\ion{O}{ii}]$\lambda\lambda7320,7330$ auroral lines are  not measured in our sample of spectra, as in \citet{2021ApJ...915...21R}, we adopted  $T_{\rm e}(\rm S^{+})$=$T_{\rm e}(\rm O^{+})$=$T_{\rm e}(\rm N^{+})$.
In objects for which it is possible to estimate directly 
$T_{\rm e}(\rm N^{+})$ (7/45) and $T_{\rm e}(\rm S^{2+})$ (14/45), these temperatures were
assumed as representative of the low and
high ionization zones, respectively. Otherwise,
when the [\ion{N}{ii}]$\lambda5755$ 
and [\ion{S}{iii}]$\lambda6312$
auroral emission-line measurements are not available, $T_{\rm e}(\rm N^{+})$ and $T_{\rm e}(\rm S^{2+})$ were derived from
the Eqs.~\ref{eq3} and \ref{eq4}, respectively.
In Table~\ref{tab2}, electron density
and temperature values for the objects in our sample are presented.

\subsection{Abundance derivation}
\label{oxyest}

For each object of our sample, using the emission line intensity ratios listed
in Table~\ref{tab1}, the electron temperature and electron
density values (listed in Table~\ref{tab2})  as well as the {\sc PyNeb} code \citep{Luridiana2015}, we derived the sulphur  ($\rm S^{+}$, $\rm S^{2+}$) and oxygen ($\rm O^{+}$, $\rm O^{2+}$) ionic abundances. 
Afterwards, applying an empirical ICF for the sulphur  and a typical value for the oxygen ICF, the total abundances for the S/H and O/H were estimated. In what follows, we describe the methodology employed in the derivation of the abundance of each considered element. 

\subsubsection{Oxygen abundance}
The total oxygen abundance in relation to the hydrogen one
 was derived assuming
 \begin{equation}
\label{eq5}
{\rm
\frac{O}{H}=ICF(O^{+}+O^{2+})\: \times \: \left[\frac{O^{+}}{H^{+}}+\frac{O^{2+}}{H^{+}}\right],} 
\end{equation}
where  ICF($\rm O^{+}+O^{2+}$) represents the Ionization Correction Factor for oxygen which takes into account the contribution of  unobservable oxygen ions, whose emission lines are observed in other spectral bands such as
X-rays (e.g. \citealt{2009A&A...505..541C, 2011A&A...530A.125C, 2010MNRAS.405..553B, 2017ApJ...848...61B, 2019ApJ...872...94M, 2020MNRAS.493.3893K}) and IR 
(e.g. \citealt{2012ApJ...746..168D, 2016ApJS..226...19F}). The $\rm O^{2+}/H^{+}$ ionic
abundance was calculated by using the [\ion{O}{iii}]$\lambda5007/\rm H\beta$ line ratio
and assuming the direct $T_{\rm e}(\rm O^{2+})$ 
and $N_{\rm e}$ values
derived from $RO3$ and $RS2$, respectively.
The $\rm O^{+}/H^{+}$ abundance was calculated  from the [\ion{O}{ii}]$\lambda3727/\rm H\beta$ emission line ratio 
and assuming $T_{\rm e}(\rm O^{+})$=$T_{\rm e}(\rm N^{+})$ with $T_{\rm e}(\rm N^{+})$ estimated from
the empirical relation given by Eq.~\ref{eq3} when the [\ion{N}{ii}]$\lambda5755$ auroral emission-line measurement is not available.

To derive ICF($\rm O^{+}+O^{2+}$) it is necessary to calculate the $\rm He^{+}/H^{+}$ and  $\rm He^{2+}/H^{+}$ ionic abundances (e.g. \citealt{1977RMxAA...2..181T, 2006A&A...448..955I, 2020MNRAS.496.2191F}), which  is not possible  because in most of the AGN spectra from our sample, the helium recombination line  $\lambda 4686\:\angstrom$ is not measured. Therefore, for consistency, the ICF($\rm O^{+}+O^{2+}$) is  assumed to have a value of 1.50 for all objects, which is an average value derived  by  \citet{2022MNRAS.514.5506D}, who found ICF  values ranging from 1.30 to 1.70 for a sample of 65 local ($z\: \la \: 0.2$) Sy~2s. 
This ICF value
translates into an abundance correction
of $\sim0.2$ dex, i.e., somewhat higher than the uncertainty ($\sim0.1$ dex) of abundances usually relied on for the $T_{\rm e}$-method
(e.g. \citealt{2003ApJ...591..801K, 2008MNRAS.383..209H}).

\subsubsection{Sulphur  abundance}
\label{subs}

The $\rm S^{+}/H^{+}$ ionic abundance for each object of our sample was derived by using the [\ion{S}{ii}]$(\lambda6716+\lambda6731)/\rm H\beta$ line intensities ratio
and assuming  $T_{\rm e}$($\rm S^{+}$)=$T_{\rm e}$($\rm N^{+}$), where  $T_{\rm e}$($\rm N^{+}$) was calculated from Eq.~\ref{eq3} when the [\ion{N}{ii}]$\lambda5755$ auroral emission-line measurement is not available.  Due to the similarity between the  ionization potentials of  $\rm S^{+}$ and $\rm N^{+}$ (23.33 eV and
29.60 eV, respectively) these ions are approximately  located in the same gas region and the use of a common temperature for both  is a good approach as largely used in SF chemical abundance studies (e.g. \citealt{2003ApJ...591..801K}). However, \citet{2021ApJ...915...21R}, who compared SF direct estimates of $T_{\rm e} (\rm S^{+})$ [derived from
$RAS2$=[\ion{S}{ii}]$(\lambda6716+\lambda6731)/(\lambda4068+\lambda4074)$] with $T_{\rm e} (\rm N^{+})$
[derived from
$RN2$=[\ion{N}{ii}]$(\lambda6548+\lambda6584)/(\lambda5755)$], found somewhat 
higher $T_{\rm e} (\rm S^{+})$ values than $T_{\rm e} (\rm N^{+})$, with an intrinsic dispersion of $\sim950$ K
between these temperatures.  Unfortunately, $RAS2$ values for Sy~2 are rarely found in the literature thus far, which makes it impossible to verify whether any of these conclusions also apply to AGNs.

Similarly, the $\rm S^{2+}/H^{+}$ was derived by using the [\ion{S}{iii}]$\lambda9069/\rm H\beta$ line intensities ratio listed in Table~\ref{tab1} and the
$T_{\rm e} (\rm S^{2+})$ values from Eq.~\ref{eq4} when the $RS3$ line ratio is not available.
The uncertainties associated with the sulphur  ionic estimates is mainly due to the error in the measurements of the emission-line fluxes and the uncertainties in the temperature values. In Table~\ref{tab3}, the sulphur  ionic abundance values for each object of the sample are listed. 

The total sulphur  abundance in relation to the hydrogen one was considered to be
\begin{equation}
\label{eq7}
{\rm
\frac{S}{H}=ICF(S^{+}+S^{2+})\: \times \: \left[\frac{S^{+}}{H^{+}}+\frac{S^{2+}}{H^{+}}\right]},  
\end{equation}  
where  ICF($\rm S^{+}+S^{2+}$) is the Ionization Correction Factor for sulphur. Since AGNs  have  harder ionizing sources than typical SFs (e.g. \citealt{2016MNRAS.456.3354F}), it is expected that  the gas phase of these objects contains  the presence of ions with  higher ionization level than $\rm S^{2+}$. In fact, 
[\ion{S}{viii}]$\lambda9910$ and [\ion{S}{ix}]$\lambda12520$ emission lines were observed in large number of AGNs in the sample presented by \citet{2006A&A...457...61R}. 
However, measurements
of  $\rm S^{3+}$ lines  (e.g. [\ion{S}{iv}]$10.51\micron$) are not available in the literature, which makes it impossible for the derivation of an empirical ICF(S) for AGNs, such as derived for SFs by  \citet{2016MNRAS.456.4407D}. 

\begin{figure}
\includegraphics[angle=0.0,width=0.47\textwidth]{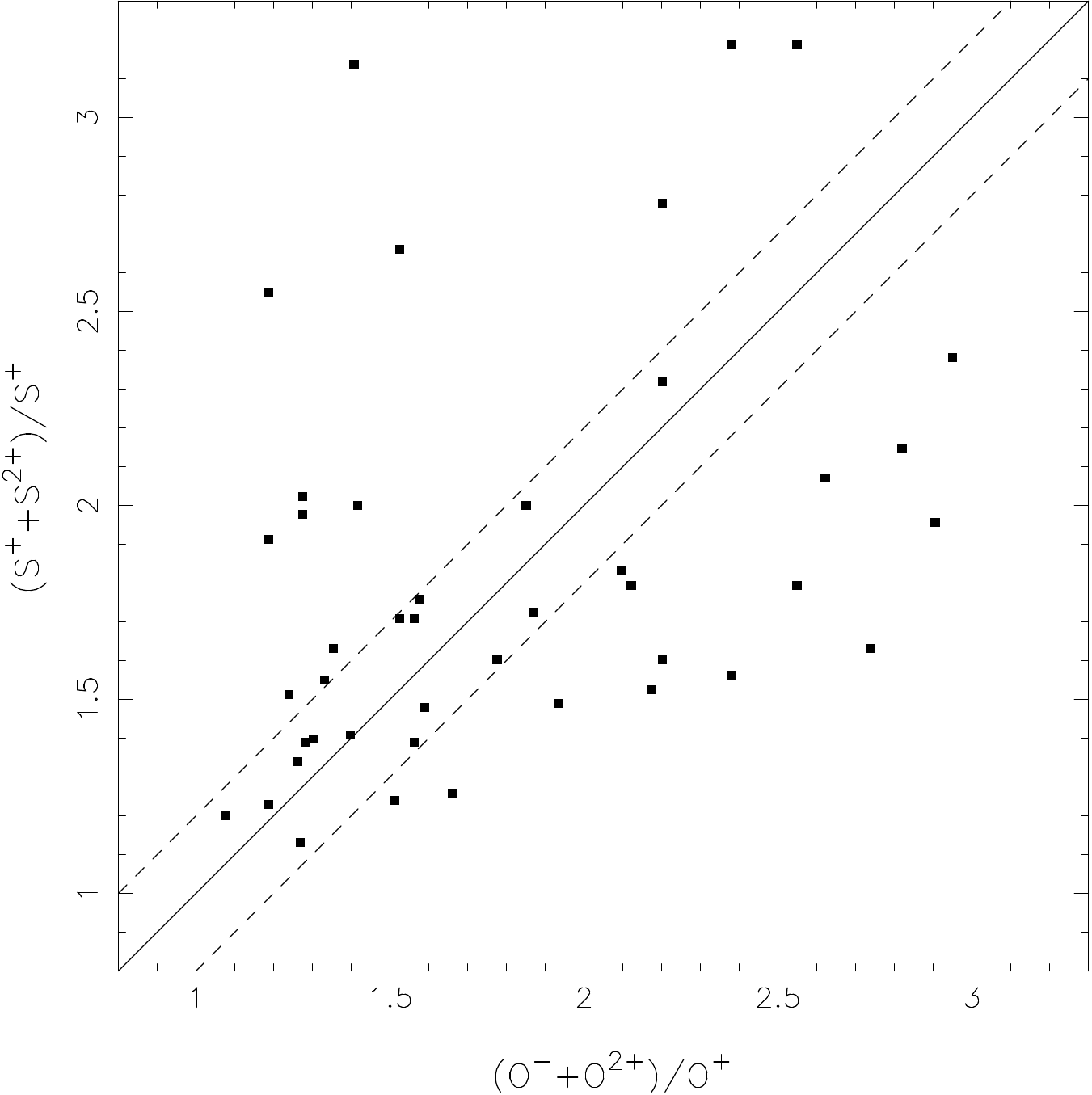}
\caption{Sulphur  versus oxygen ionic abundance ratios for our
sample of objects (see Sect.~\ref{observ}) calculated
via the $T_{\rm e}$-method (see Sect.~\ref{abund}). Solid line represents the equality between the estimates.
Dashed lines correspond to the deviation of $20\,\%$ from the one-one relation representing the mean error in the ionic abundance
estimations.}
\label{fig6}
\end{figure}

The first ICF for sulphur  (likewise for other elements) was proposed for SFs 
 by \citet{1969BOTT....5....3P} and it is given by
\begin{equation}
\label{eqicf}
 \rm ICF(S^{+}+S^{2+})=\frac{\rm S^{+}+S^{2+}}{\rm S^{+}}=\frac{\rm O^{+}+O^{2+}}{\rm O^{+}}. 
\end{equation}  
\citet{2021ApJ...915...21R} tested the application of this ICF for SFs (see also \citealt{2022MNRAS.511.4377D})
and pointed out that it is particularly reliable for low ionization degree, i.e.
when $\rm O^{+}$ zone is more dominant than the $\rm O^{2+}$ zone [$(\rm O^{+}/O) \: > \: 0.6$]. Hitherto, there has not been
sulphur  ICFs for AGNs in the literature and it
is unknown if the above relation  is completely valid for this object class. Therefore, in order to ascertain whether the  sulphur  ICF can be applied  to Sy~2s, we performed a simple test  to verify the equality indicated in Eq.~\ref{eqicf}.
The  ionic ratios for our sample are plotted in Fig.~\ref{fig6},
where the black solid line represents the equality between the estimates, while the dashed lines represent the deviations of $20\,\%$ from the one-one relation.
It can be seen that, despite the scattering, most of the ionic abundance ratio estimates  are located around of the one-one relation. Thus, we 
assumed as valid the Eq.~\ref{eqicf} for NLR
sulphur  abundance estimates.

Ionic and total oxygen and sulphur  abundance estimates derived through the $T_{\rm e}$-method together with the sulphur  ICFs for each object in our sample are listed in Table \ref{tab3}.  
 
\section{Results \& Discussion}
\label{rdisc}

\subsection{Temperature estimates}
\label{subte}
Recent studies of spatially resolved central parts
of galaxies have uncovered  the temperature structure of a few AGNs. For instance, \citet{2021ApJ...910..139R}, using Hubble Space Telescope and Apache Point Observatory spectroscopy, obtained direct estimates of $T_{\rm e}(\rm O^{2+})$ along the radius of the Sy~2 nucleus of Mrk\,78 and found temperatures in the range  10\,000-15\,000 K, with no systematic variation (see also \citealt{2018ApJ...856...46R, 2018ApJ...867...88R}). \citet{2021MNRAS.506L..11R}
used Gemini GMOS-IFU observations of three luminous nearby Seyfert galaxies (Mrk\,79, Mrk\,348 and Mrk\,607) and estimated $T_{\rm e}(\rm O^{2+}$)  fluctuations in the inner 0.4–1.1 kpc region of these galaxies. These authors found temperature fluctuations similar to those derived  in SFs and  PNe. Despite the revelations provided by these recent studies, 
an advance in the understanding of the temperature structure of AGNs, additional point-to-point or integrated $T_{\rm e}$ estimates
through distinct emission lines [e.g. [\ion{S}{III}]$(\lambda9069+\lambda9532)/\lambda6312$ and
[\ion{N}{II}]$(\lambda6548+\lambda6584)/\lambda5755$] are rare in the literature, prompting further investigation of the AGN temperature structure. Thus, our temperature estimates provide valuable knowledge to the nature of NLRs.   

\begin{figure}
\includegraphics[angle=0.0,width=0.47\textwidth]{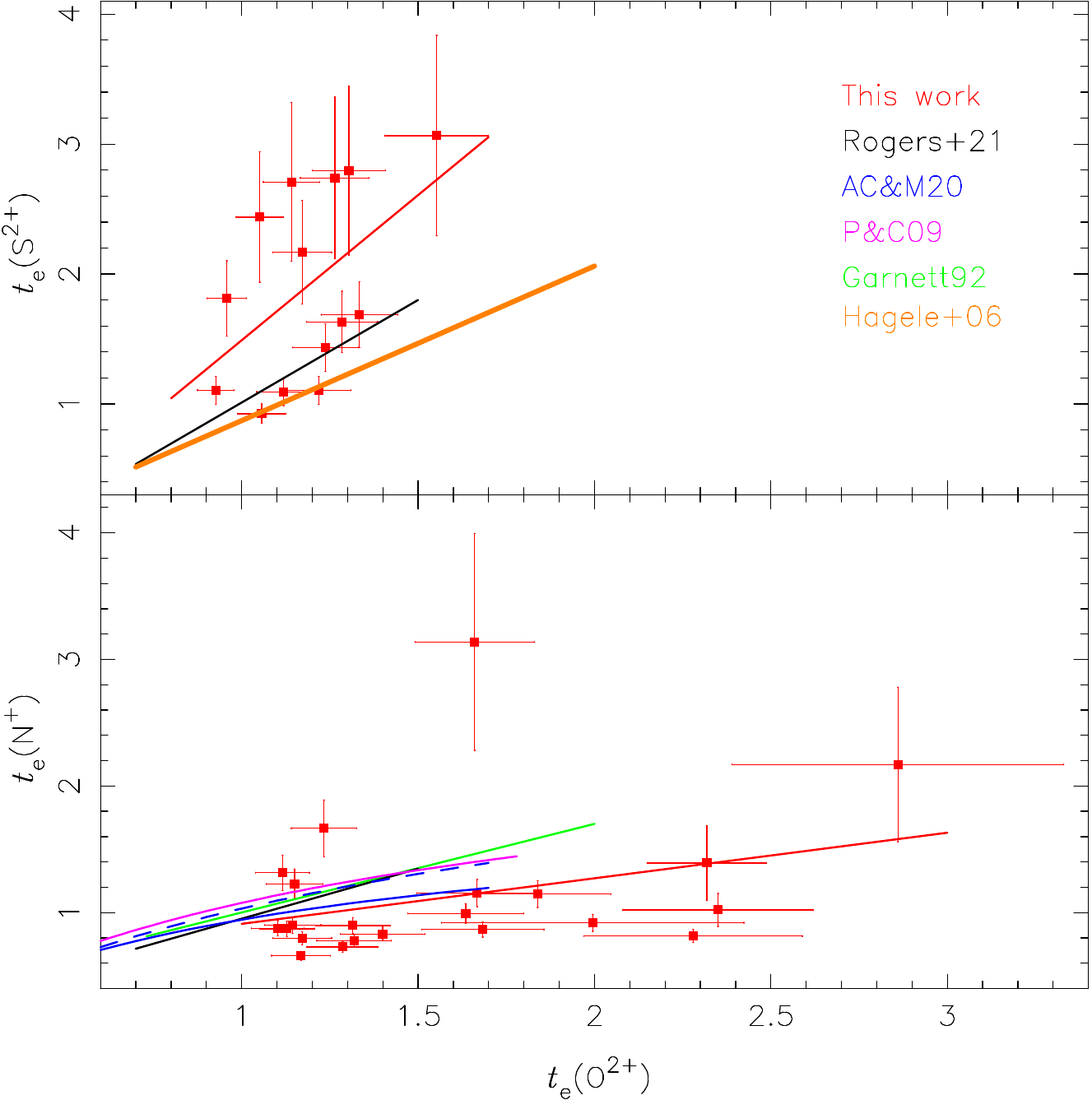}
\caption{Same as Fig.~\ref{fig5} but comparing our temperature
estimates (red points) and temperature relations (red lines
represented by Eqs.~\ref{eq3} and \ref{eq4}) with relations obtained from
SFs. Orange solid line plots the empirical relation (Eq.\ \ref{eq9}) obtained by \citet{2006MNRAS.372..293H} using direct temperature estimates for a large sample of \ion{H}{ii} galaxies, giant extragalactic \ion{H}{ii} regions, Galactic \ion{H}{ii} regions, and \ion{H}{ii} regions from the MCs. Black solid lines represent the empirical relations (Eqs.~\ref{eq10} and \ref{eq11}) based on
direct estimates of disk \ion{H}{ii} regions  by
\citet{2021ApJ...915...21R}. Blue dashed and solid lines, shown only in bottom panel, represent empirical relations for $P \: < \: 0.5$ and for $P \: > \: 0.5$, respectively, based on data compiled
from literature by \citet{2020MNRAS.497..672A}. 
P is defined as $([\ion{O}{iii}]\lambda4959+\lambda5007)/[\ion{O}{ii}]\lambda3727+[\ion{O}{iii}]\lambda4959+\lambda5007$.
Green and pink lines
represent the theoretical relations  derived by \citet[][Eq.\ref{eqgarnett}]{1992AJ....103.1330G} and \citet[][Eq.\ref{eqpm09}]{2009MNRAS.398..949P}, respectively.}
\label{fig8}
\end{figure}

A large number of AGNs present gas outflows
(e.g. \citealt{2020MNRAS.496.4857R, 2022arXiv220913125A}), shocks (e.g. \citealt{1984A&A...140..368A, 1995ApJ...455..468D,  2021MNRAS.501.1370D}) and  neutral gas reservoirs which can coexist with the ionized gas (e.g. \citealt{2014A&A...567A.125G}) and, in combination
with the hard ionization source (e.g. \citealt{2016MNRAS.456.3354F}), trend to produce a more
complex gas structure than that of SFs (e.g. \citealt{2006MNRAS.372..293H,2010MNRAS.408.2234G,2011A&A...532A.141P,2012A&A...544A..60M, 2018ApJ...867..149D}). However, complex temperature structures
can also be observed 
in SFs (e.g., \citealt{2005ApJ...619..755D, 2022ApJ...927...37J}) produced, for instance,  by starburst-driven outflows cooling (e.g. \citealt{2022ApJ...937...68D}). 
On the scenario where distinct physical processes drive the gas structure, it is expected that temperature relations
of AGNs tend to differ from the ones derived for SFs.
In order to test this hypothesis, in Fig.~\ref{fig8}, we compare
our Sy~2 temperature estimates and empirical relations 
(Eqs.~\ref{eq3} and \ref{eq4})
with those derived for SFs by the following authors:
\begin{enumerate}
 \item 
\citet{2006MNRAS.372..293H}: these authors used their own high quality spectra of \ion{H}{ii} galaxies and a large literature compilation of \ion{H}{ii} galaxies, Giant Extragalactic \ion{H}{ii} regions (GHRs), Galactic \ion{H}{ii} regions, and \ion{H}{ii} regions from the Magellanic Clouds performed by \citet{2006A&A...449..193P} to analyse the relation between $T_{\rm e}(\rm O^{2+})$ and $T_{\rm e}(\rm S^{2+})$ values derived using the $T_{\rm e}$-method (see Fig.\,7 of H\"agele and collaborators). Their linear fitting to the complete sample gave the relation: 
\begin{equation}
\label{eq9}
t_{\mathrm{e}}(\mathrm{S^{2+}})=  1.19(\pm0.08)\times t_{\mathrm{e}}(\mathrm{O^{2+}}) - {\rm 0.32(\pm0.10)},
\end{equation} 
\noindent which has a validity range of $0.70 \: \lessapprox \: t_{\mathrm{e}}(\mathrm{O^{2+}}) \: \lessapprox \: 2.0$.

    \item \citet{2021ApJ...915...21R}: the estimates by these authors include temperature values obtained for a large number of disk \ion{H}{ii} regions in the spiral galaxy NGC\,2403 using the  $T_{\rm e}$-method.
  These estimates combined with those
    of \ion{H}{ii} regions in four spiral galaxies (see Fig.~3 by \citealt{2021ApJ...915...21R}) resulted in the 
    relations:
    \begin{equation}
\label{eq10}
t_{\mathrm{e}}(\mathrm{N^{+}})=  0.79(\pm0.14)\times t_{\mathrm{e}}(\mathrm{O^{2+}}) + {\rm 0.16(\pm0.13)},
\end{equation} 
and 
 \begin{equation}
\label{eq11}
t_{\mathrm{e}}(\mathrm{S^{2+}})=  1.58(\pm0.17)\times t_{\mathrm{e}}(\mathrm{O^{2+}}) - {\rm 0.57(\pm0.16)},
\end{equation} 
which are valid for $0.70 \: \lessapprox \: t_{\mathrm{e}}(\mathrm{O^{2+}}) \: \lessapprox \: 1.5$.
\item \citet{2020MNRAS.497..672A}: these authors compiled
  emission line intensities of \ion{H}{ii} regions from the literature to explore the behaviour of the 
$T_{\rm e}(\rm N^{+})$-$T_{\rm e}(\rm O^{2+})$
temperature relation. These authors found that this relation has
a dependence on the gas ionization degree, which is traced by the line ratio  
$P=([\ion{O}{iii}]\lambda4959+\lambda5007)/[\ion{O}{ii}]\lambda3727+[\ion{O}{iii}]\lambda4959+\lambda5007$.
The following relations, which hold for $0.60 \: \lessapprox \: t_{\mathrm{e}}(\mathrm{O^{2+}}) \: \lessapprox \: 1.7$, were derived:\\
For $P \: < \: 0.5$
\begin{equation}
\label{eq12}
\frac{1}{t_{\mathrm{e}}(\mathrm{N^{+}})}=\frac{0.54(\pm0.05)}{t_{\mathrm{e}}(\mathrm{O^{2+}})}+0.52(\pm0.08)
\end{equation}
and for $P \: > \: 0.5$
\begin{equation}
\label{eq13}
\frac{1}{t_{\mathrm{e}}(\mathrm{N^{+}})}=\frac{0.61(\pm0.04)}{t_{\mathrm{e}}(\mathrm{O^{2+}})}+0.36(\pm0.04).
\end{equation}

\item \citet{1992AJ....103.1330G}: 
this author, by using the photoionization model
results derived by \citet{1982A&AS...48..299S}, proposed the relation
\begin{equation}
\label{eqgarnett}
t_{\mathrm{e}}(\mathrm{S^{2+}})=  0.83\times t_{\mathrm{e}}(\mathrm{O^{2+}}) - {\rm 0.17},
\end{equation} 
valid for $0.4 \: \lessapprox \: t_{\mathrm{e}}(\mathrm{O^{2+}}) \: \lessapprox \: 1.8.$

\item \citet{2009MNRAS.398..949P}: using photoionization model results built through the \textsc{Cloudy} code these authors deduce a relation between  $T_{\rm e}$(O$^{2+}$) and $T_{\rm e}$(N$^{+}$) given by: 
\begin{equation}
\label{eqpm09}
t_{\mathrm{e}}(\mathrm{N^{+}})=\frac{1.85}{t_{\mathrm{e}}(\mathrm{O^{2+}})^{-1}+0.72}
\end{equation}
with a valid range $0.60 \: \lessapprox \: t_{\mathrm{e}}(\mathrm{O^{2+}}) \: \lessapprox \: 1.8$.

 \end{enumerate}

In Fig.~\ref{fig8}, the above relations are compared with our
estimates and our own temperature relations.
It can be seen in the bottom panel of Fig.~\ref{fig8} that
Sy~2 nuclei present similar $T_{\rm e}(\rm N^{+})$ values
for a given $T_{\rm e}(\rm O^{2+})$
to those from SFs. Otherwise, in Fig.~\ref{fig8}, upper panel,   $T_{\rm e}(\rm S^{2+})$ NLR estimates are higher than those in SFs.
This result indicates that NLRs have a   hotter high ionization zone than the one in SFs. This is probably due to the known fact that SEDs of AGNs are harder than the ones of SFs. Moreover, gas shocks present in AGNs can produce a very distinct temperature structure than that in SFs, where shocks have a little influence. In fact,
\citet{2021MNRAS.501.1370D} built detailed composite models of photoionization and shock ionization based on the \textsc{suma} code \citep{1989ApJ...339..689V} to reproduce optical emission lines emitted by
NLRs of 244 Sy~2 nuclei. Their models predicted an abrupt increase in temperature near the shock front, reaching values of $\sim10^{5}$ K, mainly  in shock-dominated objects (see Fig.~14 by \citealt{2021MNRAS.501.1370D}). In summary, our temperature estimates support the scenario where AGNs have complex spatial
distributions of gas-temperature  and a variety of mechanisms can drive the temperature
and ionization (e.g., see \citealt{2009ApJ...698.1852B, 2016A&A...587A.138B, 2018ApJ...867..149D, 2019A&A...622A.128F}).

\begin{figure*}
\includegraphics[angle=0.0,width=0.47\textwidth]{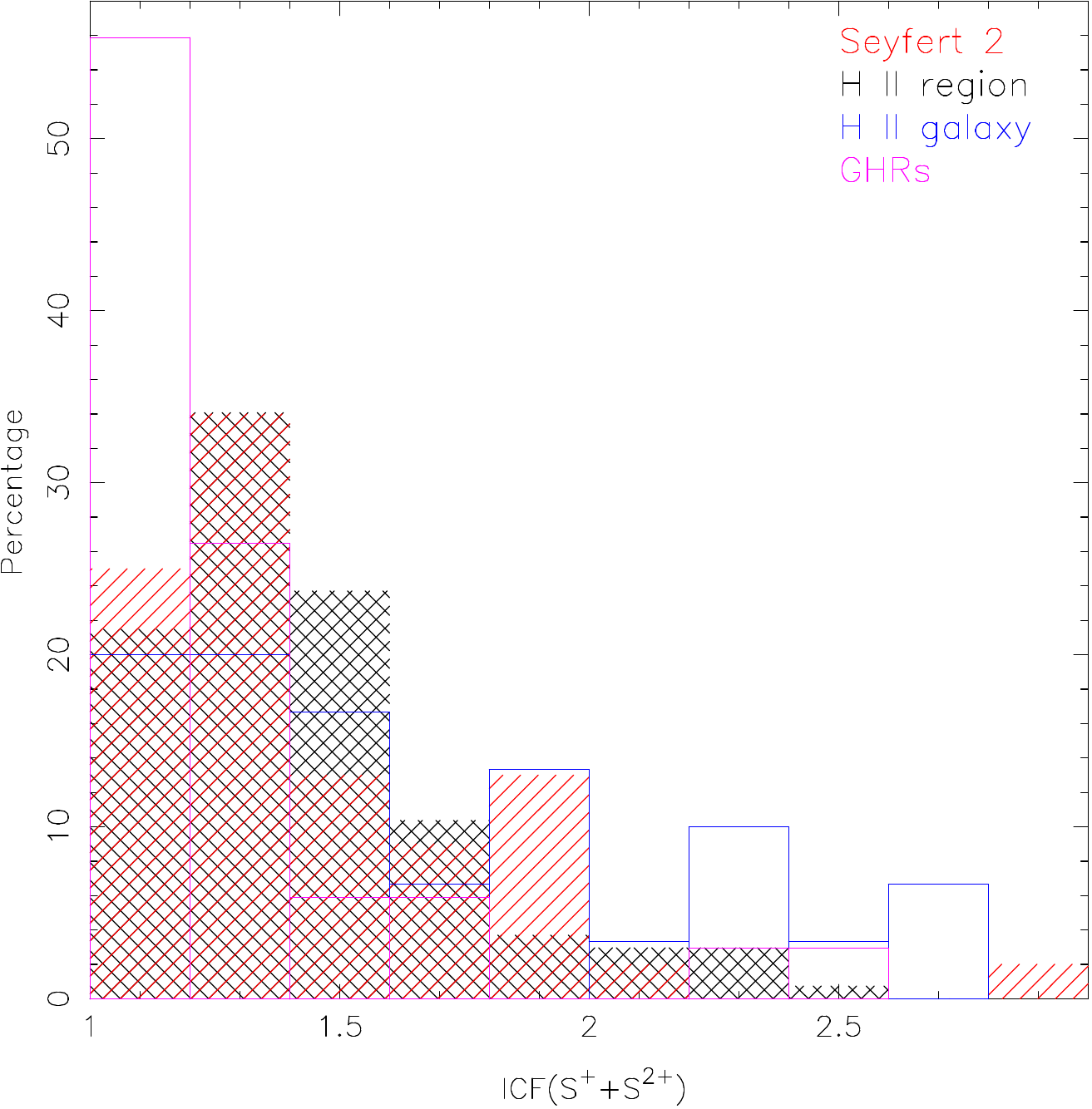}
\includegraphics[angle=0.0,width=0.47\textwidth]{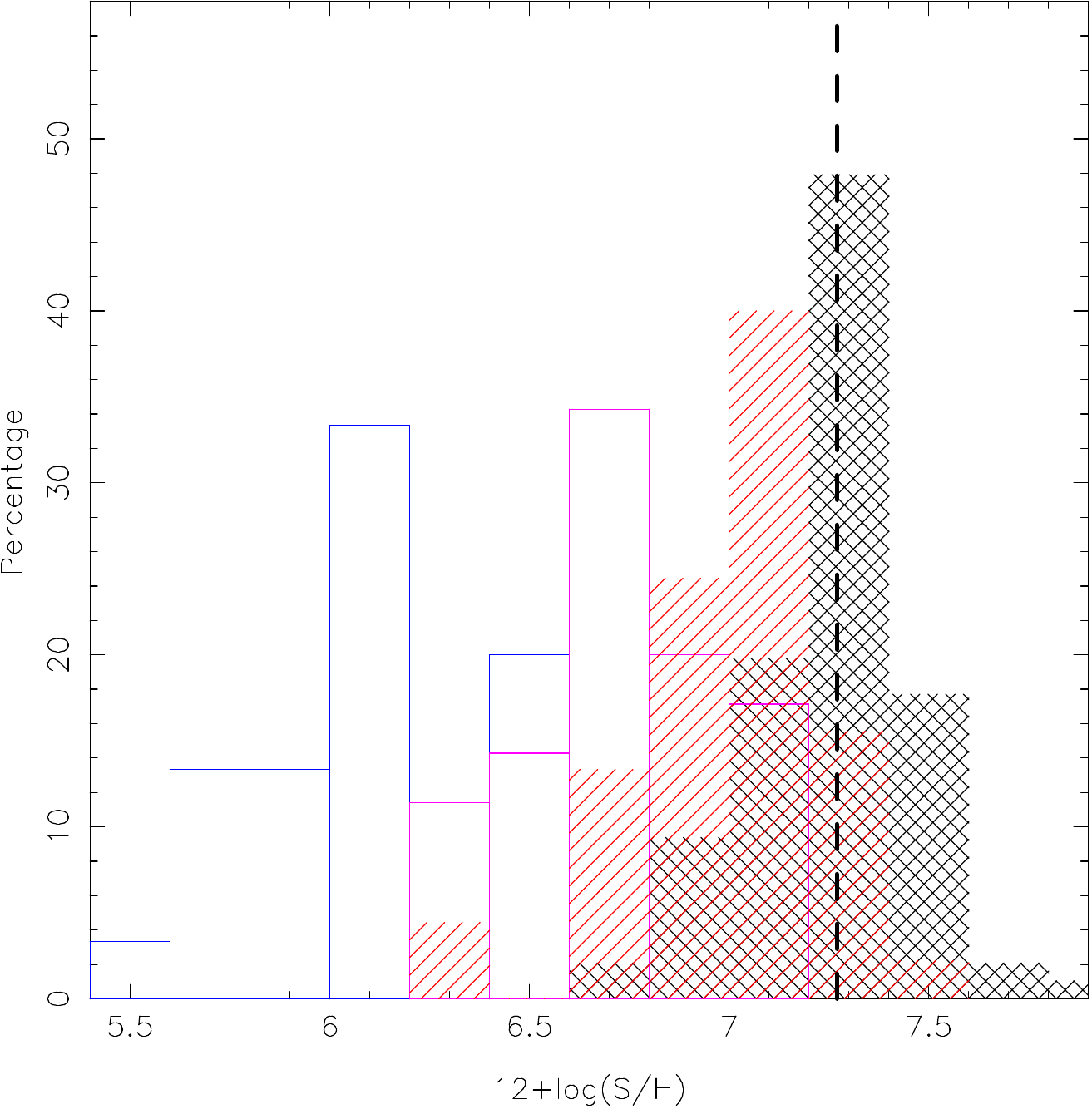}
\caption{Left panel: distributions for the sulphur  ICF (see Sect.~\ref{subs}). Red distribution is derived from our Sy~2 sample by using Eq.~\ref{eqicf} (values
listed in Table~\ref{tab3}). Black distribution is
derived by using values from disk \ion{H}{ii} region estimates
(i.e. results from CHAOS project) assuming Eq.~\ref{eqicf} for $\rm (O^{+}/O) \:  \ga \:0.6$ and the theoretical ICF from \citet{1995ApJ...445..108T} when
$\rm (O^{+}/O) \:  \la \:0.6$. Blue  distribution corresponds to \ion{H}{ii} galaxies values from \citet{2006MNRAS.372..293H,2008MNRAS.383..209H,2011MNRAS.414..272H,2012MNRAS.422.3475H} and pink distribution corresponds to GHRs from \citet{2006MNRAS.372..293H}, and for both kind of objects the approach proposed by \citet{1980ApJ...240...99B} was used. Right panel: distributions for the sulphur 
abundance.  
Dashed line represents the sulphur solar abundance
$\rm 12+\log(S/H)_{\odot}=7.27$ derived by \citet{1998SSRv...85..161G}. The color code is the same as in left panel.
}
\label{fig10}
\end{figure*}  
 
 \subsection{Sulphur  abundances}
 \label{subsulf}

According to the inside-out scheme, galaxies begin to form stars in their inner regions before the outer ones 
 (e.g.\ \citealt{1997ApJ...476..544S, 1999A&A...350..827P, 2000MNRAS.312..398B, 2003ApJ...596...47S, 2005MNRAS.358..521M, 2012ApJ...747L..28N, 2016ApJ...828...27N, 2018MNRAS.478..155V}) producing radial metallicity gradients with negative slopes (i.e. the metalicity decreases with the increase of the galactocentric distance, e.g. \citealt{2004A&A...425..849P}).
Thus, due to the location of AGNs in galactic disks, they are expected to have high abundance of heavy elements; in other words,  AGNs with low abundances
are barely found in the local universe (e.g.\ \citealt{2006MNRAS.371.1559G, 2008ApJ...687..133I, 2010A&A...517A..90I,  2017ApJ...842...44K, 2020MNRAS.492..468D}).
However,  AGNs seem to have  a more complex cosmic chemical evolution than SFs. For instance, \citet{1993ApJ...419..485M},
by means of self-consistent models of galaxy evolution, showed
that AGNs in galaxies around the lifetime of $10^{9}$ years ($z\sim 5$) reach an abundance of elements divided into two classes: ($i$) elements
with 2-3 times the solar abundance (C, Ne, O and Mg) and ($ii$) the ones
with abundances ranging from 5 to 10 times the solar abundance
(N, Ni and Fe). Conversely, low abundance or metallicity
have been derived at high redshift from SFs (see \citealt{2022MNRAS.tmp.2550C} and references therein).
Despite the fact that \citet{1993ApJ...419..485M} did not consider sulphur  (an $\alpha$-element),  its abundance would increase similarly to the oxygen abundance, i.e. the expectation will be a constant S/O abundance ratio.

Optical  surveys, such as SDSS, have
made plenty of AGNs spectroscopic data available, which make  the determinations of quantitative sulphur  abundance possible in this class of objects.  In this sense, we present a detailed analysis
of sulphur  abundance from our Sy~2 sample and a comparison with some previous results obtained
from SFs. In view of that, we consider the following SF estimates which relied on the $T_{\rm e}$-method:
\begin{itemize}
    \item \citet{2006MNRAS.372..293H}:  these authors presented
    calculations for several heavy
    elements from a sample of  \ion{H}{ii} galaxies (33 objects)  and Giant \ion{H}{ii} Regions (34 objects)  by using the $T_{\rm e}$-method. In particular, for the sulphur  abundance, they  assumed the   ICF approach proposed  by \citet{1980ApJ...240...99B} with the $\alpha$ exponent equal to 2.5. Following the same procedure,  \citet{2008MNRAS.383..209H,2011MNRAS.414..272H,2012MNRAS.422.3475H} studied another 11 \ion{H}{ii} galaxies and knots belonging to this kind of objects that we include in our control sample. 
    \item CHAOS project: the Chemical Abundances Of Spirals (\textsc{CHAOS}\footnote{\url{https://www.danielleaberg.com/chaos}}, \citealt{2015ApJ...806...16B})
     combines the power of the Large Binocular Telescope (LBT) with the broad spectral range and sensitivity of the Multi Object Double Spectrograph (MODS) to derived abundances, which relied on the $T_{\rm e}$-method, for a large sample of \ion{H}{ii} regions in spiral galaxies. Taking  these valuable data into account, we consider abundance estimates for 135 disk \ion{H}{ii} regions located in NGC\,5457, NGC\,3184 and NGC\,2403 by 
\citet{2016ApJ...830....4C}, \citet{2020ApJ...893...96B} and \citet{2021ApJ...915...21R}, respectively.  These authors adopted the sulphur  ICF given by Eq.~\ref{eqicf} for $\rm (O^{+}/O) \:  \ga \:0.6$ and the theoretical ICF from \citet{1995ApJ...445..108T} when
$\rm (O^{+}/O) \:  \la \:0.6$.
\end{itemize}
Along this section, we have used the SF abundance estimates from the above authors as benchmark. We emphasize that any selection effect, such as that which may arise as a result of  the existence of auroral lines in spectra, will be present in both our AGN sample and the SF sample taken from the literature.

 Our sulphur  ICFs for the Seyfert galaxies (listed in Table~\ref{tab3}) indicate values
between $\sim1.1$ and $\sim3.0$, with an averaged value
of $1.44$, i.e. about $45\%$ of the sulphur  is in  higher ionization stages than $\rm S^{2+}$.  We found an ICF value higher than 2.0 for only  two objects: 2.02 and 2.94 for Mrk\,573 and NGC\,7674, respectively.  Interestingly, for the most extreme ICF value, i.e. for NGC\,7674, \citet{2017NatAs...1..727K}, by using radio long baseline interferometry, hinted
 that this object hosts a Binary Supermassive Black Hole (for a different conclusion see \citealt{2022ApJ...933..143B}). Moreover,  additional evidence that this object has a hard ionizing spectra is the presence of emission lines
of high ionization ions [\ion{Ne}{v}]$\lambda3346,\lambda3426$, 
as observed by \citet{1994ApJ...435..171K}. 
In any case, even if this object is ruled out from the average ICF calculations, a similar value ($1.40$)  is obtained. In order to compare our sulphur  ICF values with those from SFs, we
consider the ICFs derived by   \citet{2006MNRAS.372..293H} and from the  CHAOS project. In the left panel of Fig.~\ref{fig10}, the distribution of sulphur  ICFs for Sy~2s and SFs are shown, where it can be seen that  a good agreement exist among them, even though the GHRs present a distribution peak at lower values. It is worth to be noted that most 
 of the objects ($\sim 90\%$) belonging to the distinct object classes present sulphur  ICFs lower than
$\sim 2$. The range and average of the ICFs for our sample of Sy~2, \ion{H}{ii} galaxies, GHRs and disk \ion{H}{ii} regions  are presented in Table~\ref{tabf1} showing that the different samples have similar ICF values.  Therefore, despite the fact that Sy~2s have a harder ionizing source than SFs, these two distinct object classes have similar sulphur  ionic fractions.

\begin{table*}
\caption{Range and average sulphur  ICF, S/H, O/H, and S/O abundance values for our sample of Sy~2s, disk \ion{H}{ii} regions, \ion{H}{ii} galaxies, and Giant 
\ion{H}{ii} regions.}
\label{tabf1}
\begin{tabular}{@{}lcccccccccccc@{}}	 
\noalign{\smallskip} 
\hline 
                           &\multicolumn{2}{c}{ICF($\rm S^{+}+S^{2+}$)} &  & \multicolumn{2}{c}{12+log(S/H)} &  & \multicolumn{2}{c}{12+log(O/H)} &  &  \multicolumn{2}{c}{log(S/O)}       & Ref.\\

\cline{2-3}
\cline{5-6}
\cline{8-9}
\cline{11-12}
Object type                & Range       &      Average       &  &    Range    & Average   	             &  & Range        &  Average	  &  &	   Range	   & Average	    &	\\    
\hline          
Sy~2                  &  1.1 - 3.0  &    $1.44$   &  &  6.2 - 7.5  &  $6.98\pm0.25$              &  & 8.0 - 9.1    &  $8.71\pm0.24$	  &  &    $-2.4$ - $-1.3$  & $-1.64\pm0.20$ &  1  \\   
\ion{H}{ii} region         &  1.1 - 2.5  &    $1.43$   &  &  6.2 - 7.9  &  $7.11\pm0.28$	             &  & 7.8 - 8.9    &  $8.47\pm0.19$   &  &    $-1.7$ - $-0.9$  & $-1.35\pm0.15$ &  2  \\   
\ion{H}{ii} galaxy         &  1.0 - 2.8  &    $1.66$   &  &  5.5 - 6.6  &  $6.10\pm0.27$              &  & 7.0 - 8.2    &  $7.79\pm0.30$	  &  &    $-2.0$ - $-1.4$  & $-1.67\pm0.15$ &  3,4  \\   
GHR                       &  1.0 - 2.5  &    $1.26$   &  &  6.2 - 7.2  &  $6.73\pm0.24$	             &  & 7.6 - 8.6    &  $8.12\pm0.23$	  &  &    $-1.9$ - $-0.7$  & $-1.38\pm0.30$ &  4  \\   
\hline            
\end{tabular}	   									
\begin{minipage}[c]{2\columnwidth}
(1) This work. (2) \citet{2016ApJ...830....4C}, \citet{2020ApJ...893...96B} and \citet{2021ApJ...915...21R}. (3) \cite{2008MNRAS.383..209H, 2011MNRAS.414..272H,2012MNRAS.422.3475H}. (4) \citet{2006MNRAS.372..293H}.
\end{minipage} 
\end{table*}

Concerning the total sulphur  abundance, in the right panel of Fig.~\ref{fig10}, we present the S/H abundance distribution for our sample of Sy~2 nuclei, \ion{H}{ii} galaxies and GHRs from \citet{2006MNRAS.372..293H,2008MNRAS.383..209H,2011MNRAS.414..272H,2012MNRAS.422.3475H} and for disk \ion{H}{ii} regions from the \textsc{CHAOS} project.
Also in this plot, the sulphur solar abundance
$\rm 12+\log(S/H)_{\odot}=7.27$ derived by \citet{1998SSRv...85..161G} is represented by the dashed line.
We note that the Sy~2s
present an intermediate
S/H distribution between that of
GHRs and the disk \ion{H}{ii} regions, while \ion{H}{ii} galaxies  tend to present lower sulphur  abundances.  In Table~\ref{tabf1} the range and S/H average values for our AGN sample and for the SF benchmark sample are listed. The Sy~2s present  S/H values in the range of $6.2 \: < \: 12+\log(\rm S/H) \: < \: 7.5$  and considering the sulphur  solar value as
12+$\log(\rm S/H)_{\odot}=7.26$ \citep{1998SSRv...85..161G}, represents
$0.1 \: < \: (\rm S/S_{\odot}) \: < \: 1.8$, where most of the objects (40/45)  have subsolar sulphur  abundance.  This result can be biased 
due to the fact that we  selected only objects which have the 
[\ion{S}{iii}]$\lambda9069$ and auroral line [\ion{O}{iii}]$\lambda4363$  measured, 
resulting (see Table~\ref{tabf1})
only in Sy~2s with O/H values in the range $8.0 \: \la \: 12+\log(\rm O/H) \: \la \: 9.1$ or $0.2 \: \la \: (Z/\rm Z_{\odot}) \: \la \: 2.6$, adopting the solar oxygen  value of
$12+\log(\rm O/H)_{\odot}=8.69$ \citep{2001ApJ...556L..63A}. \citet{2020MNRAS.492..468D}, who considered a sample of 463 confirmed Seyfert 2 AGNs ($z \: \la \: 0.4$) and  used distinct methods which did not necessarily required auroral lines, found values in the range
$7.1 \: \la \: 12+\log(\rm O/H) \: \la \: 9.2$ or $0.02 \: \la \: 
(Z/\rm Z_{\odot}) \: \la \: 3.2$. Therefore,  lower and higher S/H abundances would be probably derived in the sample if the data by \citet{2020MNRAS.492..468D} could be  taken into  account. In any case, the abundance estimates from the \textsc{CHAOS} project comprises  inner disk \ion{H}{ii} regions, therefore, it is expected that these objects and Sy~2 nuclei would have similar    S/H abundances, when a large sample of objects is considered. Interestingly, the maximum S/H value ($\sim 7.9$ dex)
is derived for disk \ion{H}{ii} regions while sulphur  abundances in Sy~2s reach up
to $\sim 7.5$ dex.
This result points to a distinct   chemical enrichment of the ISM  near the  AGNs in comparison to that of the innermost disk \ion{H}{ii} regions.
We emphasize that a more detailed comparison taking into account  SFs and AGNs located in galaxies with similar mass (see \citealt{2022MNRAS.513..807D})  is need to confirm this result.

\begin{figure}
\includegraphics[angle=0.0,width=0.47\textwidth]{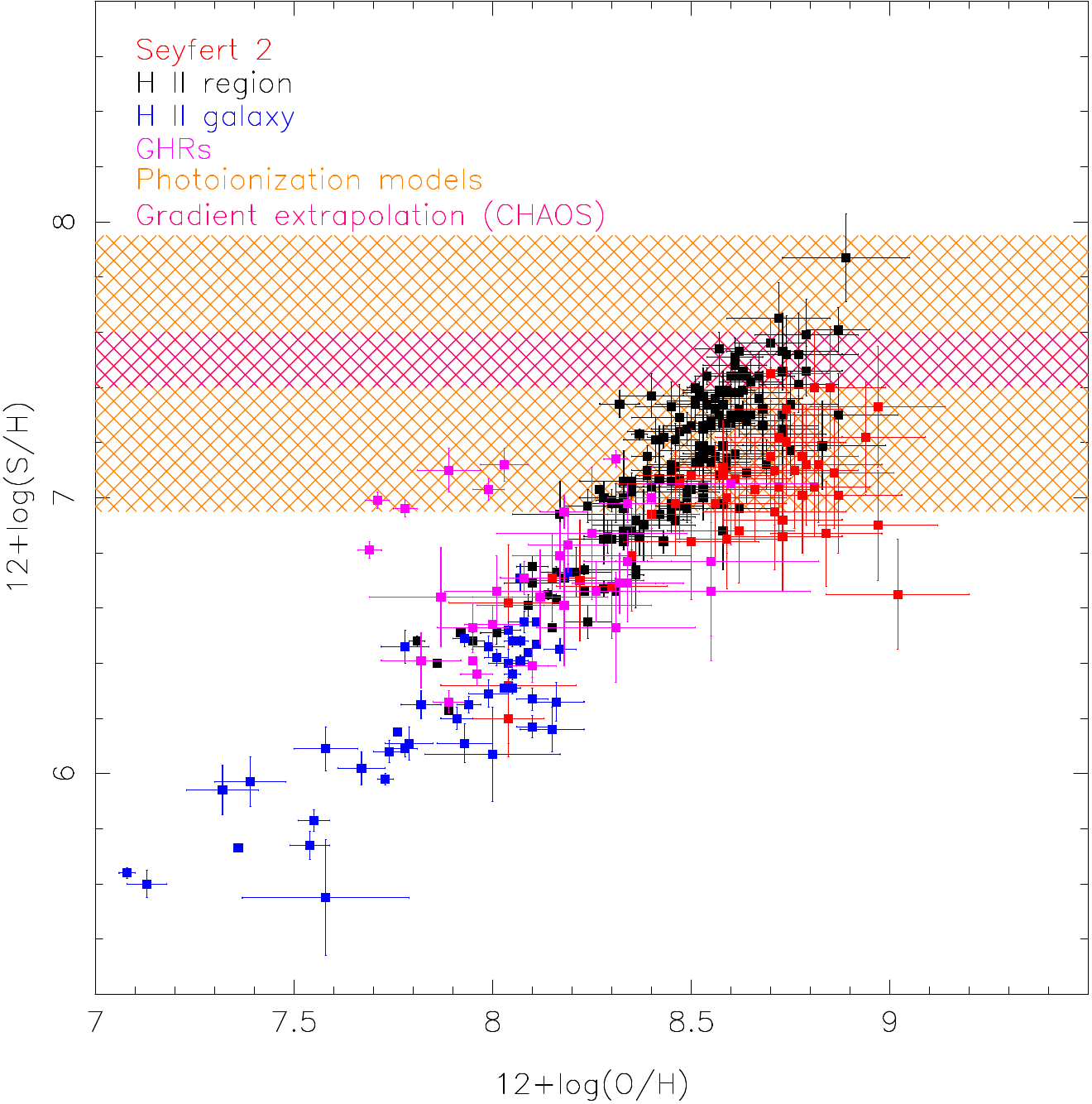}
\caption{Relation between the total sulphur  and oxygen abundances [12+log(S/H) vs. 12+log(O/H)].
Red points represent our Sy~2 nuclei direct estimates. 
Black points represent estimates from disk \ion{H}{ii} regions also obtained  through the $T_{\rm e}$-method from the CHAOS project. Pink points represent GHRs estimated from \citet{2006MNRAS.372..293H} and blue ones those estimates for \ion{H}{ii} galaxies from \citet{2006MNRAS.372..293H,2008MNRAS.383..209H,2011MNRAS.414..272H,2012MNRAS.422.3475H}.
The orange hatched area represents the range of S/H values
inferred through photoionization models for a
sample of 177 Seyfert galaxies by \citet{thaisa90}.
The magenta hatched area represents the range of S/H values
obtained from the extrapolation of radial abundance gradients to central parts of four (NGC\,0628, NGC\,5194, NGC\,5457, NGC\,3184) spiral galaxies by
\citet{2020ApJ...893...96B}.}
\label{fig11}
\end{figure}

\begin{figure}
\includegraphics[angle=0.0,width=0.47\textwidth]{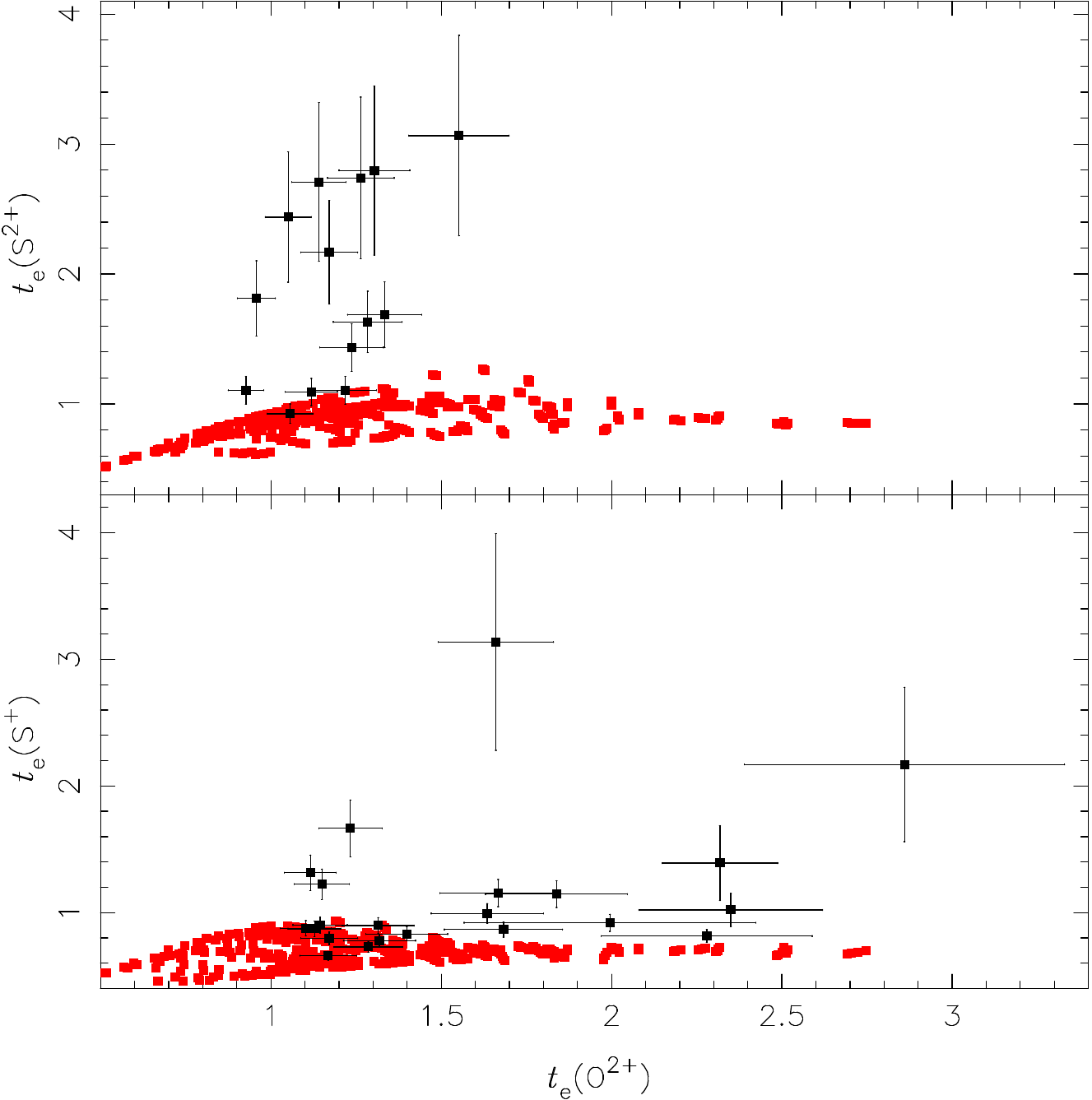}
\caption{Temperatures (in units of $10^{4}$ K) for the $\rm S^{+}$ (bottom panel) and  
$\rm S^{2+}$ (upper panel) versus the temperature for $\rm O^{2+}$.
Black points are Sy~2 direct estimations which relied on auroral lines (see Sect.~\ref{subte}).
Red points are temperature predictions by
AGN photoionization models built with the \textsc{Cloudy} code by \citet{2020MNRAS.492.5675C}.}
\label{fig12}
\end{figure}

In Fig.~\ref{fig11}, we show a plot of the S/H versus O/H abundances for our Sy~2 sample and for the SF benchmark. Also in this figure, the range of the S/H values derived from the photoionization models by \citet{thaisa90} and considering a larger sample of AGNs  than 
our data, is indicated.  \citet{2020ApJ...893...96B} summarized the radial sulphur  gradients (and the gradients for other elements) in four spiral galaxies from the \textsc{CHAOS} project, which are represented by \begin{equation}
    12+\log(\mathrm{S/H})= grad(\mathrm{S}) \times R + Y_{0}(\mathrm{S}), 
\end{equation}
where $grad(\rm S)$ is the value of the slope of the sulphur  gradient, $R$ is the radial galactic distance and $Y_{0}(\rm S)$ is the extrapolated value of S/H gradient to the
galactic center $R=0$. The range of $Y_{0}(\rm S)$ derived by \citet{2020ApJ...893...96B} is also represented 
in Fig.~\ref{fig11}. We note that, for a given
O/H value, in general Sy~2s present lower S/H values than  the majority of
disk \ion{H}{ii} regions and than those from extrapolated gradients. Again, this discrepancy can be due to the distinct chemical evolution of AGNs and SFs or even
due to the small AGN sample  (the sample contains only 45 objects). \ion{H}{ii} galaxies 
 and GHRs present
 lower S/H and O/H abundances in comparison with the Sy~2s, which indicate that the former objects are less chemically evolved than the latter. Finally, the model results by \citet{thaisa90} predicted, on average, higher ($\sim0.3$ dex) S/H values than those derived by using the $T_{\rm e}$-method for Sy~2s. This discrepancy can be partly due to the known problem
of photoionization models overestimating abundances in comparison to the $T_{\rm e}$-method. In fact, \citet{2020MNRAS.496.3209D} showed that direct temperature estimates of $T_{\rm e}(\rm O^{2+})$ are higher (up to $11\:000$ K) than those predicted by photoionization models, which translates into  an overestimate of the O/H abundance of up to $\sim1$ dex (with an average value of $\sim 0.2$ dex) by the photoionization models. In order to ascertain if the temperature problem also exists in the sulphur  temperatures, in Fig.~\ref{fig12}, we compare our direct temperature estimates
(shown in Fig.~\ref{fig4}) with temperature predictions by
the photoionization models  built
with the \textsc{Cloudy} code by \citet{2020MNRAS.492.5675C}
taking into account a wide range of NLR nebular parameters:
\begin{itemize}
     \item Metallicity: ($Z/\rm Z_{\odot})=3.0, 2.0, 1.0, 0.75, 0.5$, and $0.2$.
     \item Electron density: $N_{\rm e}\: \rm (cm^{-3})=3000, 1500, 500, 100$.
     \item Ionization parameter ($U$): $\log U$  ranging from $-1.5$ to $-3.5$, with step of 0.5 dex.
     \item Spectra Energy Distribution (SED): the SED is parametrized by the  continuum between 2 keV and 2500\AA\ \citep{1979ApJ...234L...9T} and it is 
 described by a power law with a spectral index $\alpha_{ox}$=$-$0.8, $-$1.1 and $-1.4$.
 \end{itemize}
In Fig.~\ref{fig12}, it can be seen   that similar to oxygen temperatures (see \citealt{2020MNRAS.496.3209D}) direct temperature estimates for $\rm S^{+}$ and
$\rm S^{2+}$
are higher than those predicted by AGN photoioinization models\footnote{Model temperatures values in Fig.~\ref{fig12} correspond to the mean
temperature for $\rm S^{\rm +}$  and $\rm S^{2+}$ over the nebular AGN radius times the
electron density.}.
Thus, this result explains the discrepancy between the sulphur  abundance inferred
by the photoionization models built by \citet{thaisa90} and those calculated
from our sample by using the $T_{\rm e}$-method.

\begin{figure}
\includegraphics[angle=0.0,width=0.47\textwidth]{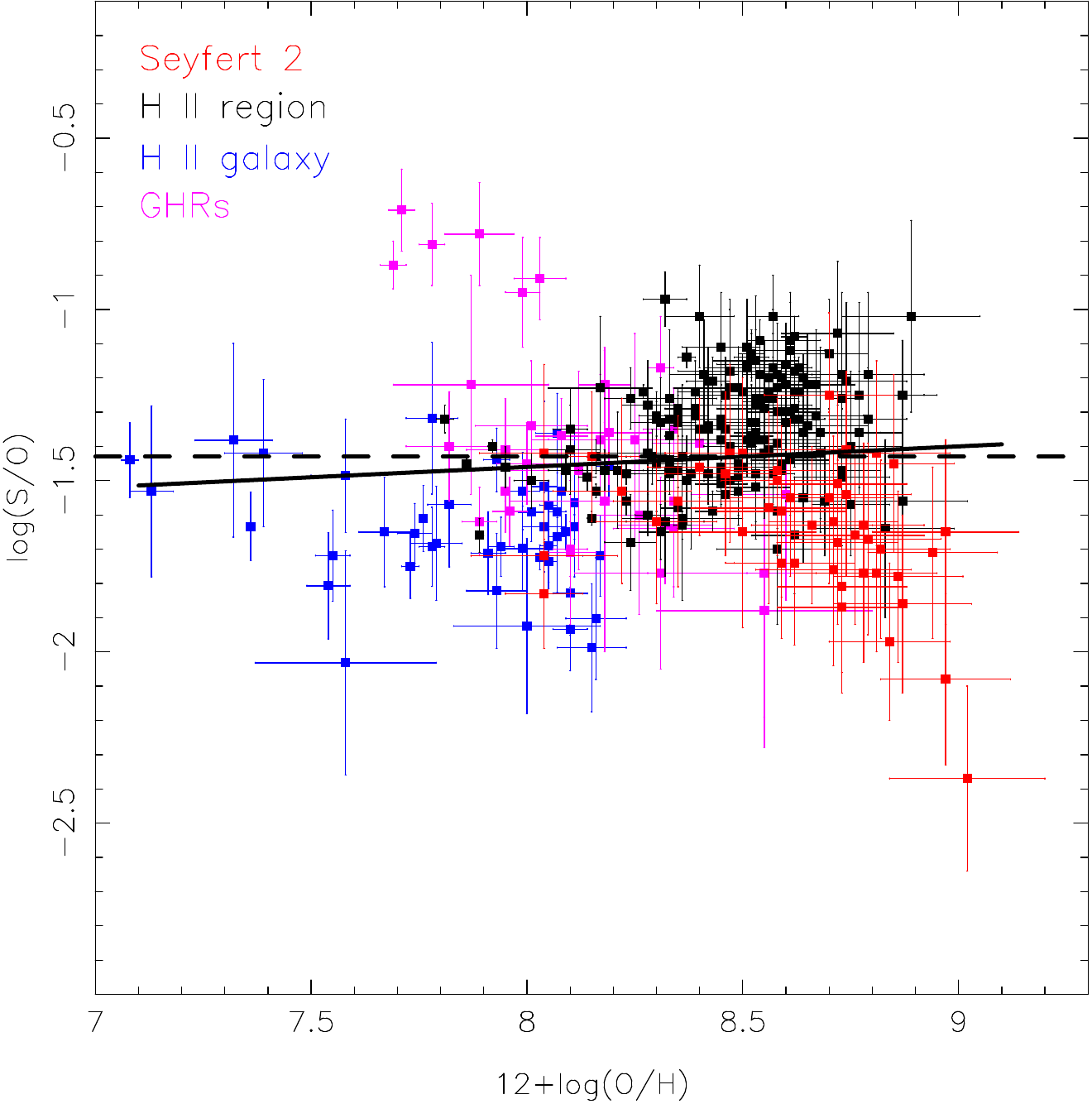}
\caption{Relation between log(S/O) and 12+log(O/H).
Red points represent our Sy~2 nuclei direct estimates. Pink points represent direct estimates for 
GHRs from \citet{2006MNRAS.372..293H}.
Black points represent estimates from disk \ion{H}{ii} regions obtained also 
through $T_{\rm e}$-method by \citet{2020ApJ...893...96B}. Blue points represent
direct estimates for \ion{H}{ii} galaxies by 
\citet{2006MNRAS.372..293H, 2008MNRAS.383..209H, 2011MNRAS.414..272H, 2012MNRAS.422.3475H}. 
Dashed line represents the  solar value of $\rm \log(S/O)_{\odot}=-1.43$. 
Solid  line   represents the linear regression considering  all the estimates given by Eq. 16.}
\label{fig13}
\end{figure}

Finally, in Fig.~\ref{fig13}, we show a plot of log(S/O) versus 12+log(O/H), which   compares our direct abundance estimates with the SF benchmark.  Considering the estimates of all objects (AGN+SF), 
we provide the following relation 
\begin{equation}
\label{finso}
 \rm \log(S/O)=+0.06(\pm0.03) \times [12+\log(O/H)] - 1.94 (\pm0.33),
\end{equation}
with the Pearson coefficient parameters ($R = 0.09\pm0.07$ and {\it p-value} = 0.24), i.e. 
there is no correlation  between the estimations.
Thus, our estimates combined with those from a large sample of
SFs suggest that  S/O is constant over a wide range of O/H, as found by  recent results from the CHAOS project  (see \citealt{2021ApJ...915...21R} and references therein). However, in Fig.~\ref{fig13}, it can be seen that there is a clear trend of
S/O values of Sy~2 decreasing with O/H in the high metallicity regime. The same behaviour was found from \ion{H}{ii} regions, for instance, by \citet{2016MNRAS.456.4407D} and \citet{2022MNRAS.511.4377D}.

It is worth to mention that  emission lines of AGNs, such as [\ion{S}{III}]$\lambda9069,\lambda9532$ and
auroral lines (mainly [\ion{N}{ii}]$\lambda5755$ and
[\ion{S}{iii}]$\lambda6312$)  are either measured with low S/N ($\sim2$)
or unavailable in the literature (e.g., see \citealt{1978ApJ...223...56K, 2015ApJS..217...12D}). This implies that
chemical abundance studies of AGNs are difficult to be carried out and, even when
it is possible to determine the
abundance directly,  higher  (a factor of $\sim2$, see Fig~\ref{fig11}) abundance errors in comparison with those
of SFs are derived. The next generation of telescopes, such as the
Giant Magellan, European Extremely Large, and Thirty Meter Telescopes, will provide higher S/N measurements of weak AGN emission lines and  will allow a breakthrough in our understanding of the chemical abundance in AGNs and  objects with very high metallicity.

\section{Conclusions}
\label{conc}

We have used observations of the intensities of narrow emission lines in the spectral interval 
 $3000 \: < \lambda($\AA$) \: < \: 9100$  of sample of 45 nearby ($z\: \la \: 0.08$) Seyfert 2 nuclei
taken from  SDSS DR17 and other compilations from the literature  to  perform direct estimations of electron temperatures through the $T_{\rm e}$-method and estimates of  sulphur  and oxygen abundances relative to hydrogen.
These estimates were compared with those from local star-forming regions, i.e. disk \ion{H}{ii} regions, \ion{H}{ii} galaxies and Giant \ion{H}{ii} Regions, whose  abundance estimates were compiled from the literature.
Regarding the electron temperatures, we found 
that Seyfert~2 and star-forming regions have 
similar temperature in the  gas regions where most $\rm S\rm ^{+}$ is located. However, this result is not derived for the zones where most $\rm S\rm ^{2+}$ is located: $\rm S^{2+}$ electron temperatures are higher ($\sim10\:000$ K) from Seyfert~2 than from  star-forming regions.
We interpret this result as, probably, due to the known feature of SEDs of AGNs are harder
than that  of SFs, producing a hotter  gas in the innermost narrow line region of AGNs.
For our sample of Seyfert~2, we derived total sulphur  abundances in the range of $6.2 \: \la 12+\log(\rm S/H) \: \la \: 7.5$ or $0.1\: \la \: (\rm S/S_{\odot}) \: \la \: 1.8$. The Seyfert~2 sulphur  abundances are lower by a factor of $\sim0.4$ dex  than those derived for SFs with similar metallicities. 
 This discrepancy can be
interpreted  as  due to a distinct chemical enrichment of the ISM near the AGNs
in comparison to that of the SFs.  
The relation between S/O and O/H abundance ratios
 derived from our  Seyfert~2 nuclei sample presents an abrupt ($\sim 0.5$ dex)
 decrease with the increase of O/H for the high metallicity regime [$\rm 12+\log(O/H) \: \ga 8.7)$], which is not derived from star-forming regions. However, when our Sy~2 estimates are combined with those  from a  large sample of star-forming regions, we did not
 find any dependence between S/O and O/H, supporting the idea that sulphur  and oxygen 
 are produced by stars with similar mass range  and  that the Initial Mass Function is universal.

\section*{Acknowledgements}

 OLD is grateful to Fundação de Amparo à
Pesquisa do Estado de São Paulo (FAPESP) and Conselho Nacional
de Desenvolvimento Científico e Tecnológico (CNPq). ACK thanks  FAPESP for the support grant 	2020/16416-5 and the Conselho Nacional de Desenvolvimento Científico e Tecnológico (CNPq). RAR acknowledges financial support from CNPq  and Funda\c c\~ao de Amparo \`a pesquisa do Estado do Rio Grande do Sul (FAPERGS). MA gratefully acknowledges support from Coordenação de Aperfeiçoamento de Pessoal de Nível Superior (CAPES). 
Funding for the Sloan Digital Sky Survey IV has been provided by the
Alfred P. Sloan Foundation, the U.S. Department of Energy Office of Science,
and the Participating Institutions. SDSS acknowledges support and resources
from the Center for High-Performance Computing at the University of Utah.
The SDSS web site is \url{www.sdss.org}.
SDSS is managed by the Astrophysical Research Consortium for the
Participating Institutions of the SDSS Collaboration including the Brazilian
Participation Group, the Carnegie Institution for Science, Carnegie Mellon
University, the Chilean Participation Group, the French Participation Group,
Harvard-Smithsonian Center for Astrophysics, Instituto de Astrofísica de
Canarias, The Johns Hopkins University, Kavli Institute for the Physics and
Mathematics of the Universe (IPMU)/University of Tokyo, the Korean
Participation Group, Lawrence Berkeley National Laboratory, Leibniz Institut
fur Astrophysik Potsdam (AIP), Max-Planck-Institut fur Astronomie (MPIA 
Heidelberg), Max-Planck-Institut fur Astrophysik (MPA Garching), 
Max-Planck-Institut fur Extraterrestrische Physik (MPE), National Astronomical 
Observatories of China, New Mexico State University, New York University,
University of Notre Dame, Observatorio Nacional/MCTI, The Ohio State
University, Pennsylvania State University, Shanghai Astronomical Observatory,
United Kingdom Participation Group, Universidad Nacional Autonoma de
Mexico, University of Arizona, University of Colorado Boulder, University of 
Oxford, University of Portsmouth, University of Utah, University of Virginia,
University of Washington, University of Wisconsin, Vanderbilt University, and
Yale University. MV acknowledges support from the CONACYT grant from the program ``Estancias Posdoctorales por M\'exico 2022''.

\section*{Data Availability}

The data underlying this article will be shared on reasonable request
to the corresponding author.

\bibliographystyle{mnras}
\bibliography{refs}

\appendix

\section{Tables}

\begin{table*}
\addtolength{\tabcolsep}{-3pt}
\caption{Emission-line intensities of Seyfert~2s relative to H$\beta$=1.00 taken from SDSS DR17 and compiled from the literature.
For Mrk\,3 the sum  [\ion{S}{ii}]$\lambda$6717+$6731$ (in relation to H$\beta$) is listed. Last column lists the references for the optical and near infrared lines, except for Mrk\,3
and ESO\,138\,G1 which the line set was taken from only
one work.}
\label{tab1}
\begin{tabular}{@{}lccccccccccl@{}}
\hline
Object         & [\ion{O}{ii}]  & [\ion{O}{iii}] & [\ion{O}{iii}] & [\ion{N}{ii}] &
[\ion{S}{iii}] & $\rm H\alpha$  & [\ion{N}{ii}]  & [\ion{S}{ii}]  &
[\ion{S}{ii}]  & [\ion{S}{iii}] & Ref. \\
\noalign{\smallskip}
 & $\lambda$3727 & $\lambda$4363 & $\lambda$5007 & $\lambda$5755 &
   $\lambda$6312 & $\lambda$6563 & $\lambda$6584 &
   $\lambda$6716 & $\lambda$6731 & $\lambda$9069+$\lambda$9532 & \\
\hline
Mrk\,573                  &  $2.13\pm0.05$		  &   $0.14\pm0.011$             &     $10.26\pm0.05$	&		       &       ---		 & $2.86\pm0.05$     &    $2.34\pm0.04$ 	  &	$0.79\pm0.01$	       &      $0.73\pm0.02$	    &	      2.58		   &	1, 12$^{(\rm a)}$ \\		     
NGC\,5728                 &  $2.32\pm0.03$		  &  $0.11\pm0.016$		 &     $9.10\pm0.11$ 	&		       &       ---		 & $2.86\pm0.13$     &    $2.86\pm0.13$ 	  &	$0.97\pm0.06$	       &      $0.66\pm0.03$	    &	      1.92		   &	1, 12$^{(\rm a)}$ \\		     
ESO\,428-G14              &  2.22  		          &   0.24  		         &     11.2		&		       &       ---		 & 2.90 	     &    3.13  		  &	0.83		       &      0.88		    &	      1.87		   &	2, 12$^{(\rm a)}$ \\		     
NGC\,4388                 &  2.73  		          &   0.16  		         &     10.55		&		       &       ---		 & 2.84 	     &    1.50  		  &	0.72		       &      0.62		    &	      1.10		   &	3, 13$^{(\rm a)}$ \\		     
Mrk\,78                   &  4.96  		          &   0.14  		         &     11.94		&		       &       ---		 & 2.46 	     &    2.46  		  &	0.68		       &      0.61		    &	      0.89		   &	4, 14$^{(\rm a)}$ \\		     
NGC\,7674                 &  1.29  		          &   0.12  		         &     12.55		&		       &       ---		 & 3.70 	     &    3.68  		  &	0.54		       &      0.64		    &	      3.15		   &	5, 12$^{(\rm a)}$ \\		     
NGC\,2110                 &  4.38  		          &   0.17  		         &      4.76		&		       &       ---		 & 2.66 	     &    3.76  		  &	1.52		       &      1.42		    &	      1.91		   &	6, 12$^{(\rm a)}$ \\		     
NGC\,7682                 &  2.85  		          &  0.16			 &     9.34		&		       &       ---		 & 3.10 	     &    3.03  		  &	1.09		       &      1.17		    &	      1.95		   &	7, 12$^{(\rm a)}$ \\		     
NGC\,3227                 &  3.22  		          &  0.50			 &    10.73		&		       &       ---		 & 2.86 	     &    5.01  		  &	1.24		       &      1.26		    &	      1.66		   &	8, 12$^{(\rm b)}$ \\		     
Mrk\,1066                 &  3.34  		          &  0.08			 &     3.84		&		       &       ---		 & 2.76 	     &    2.42  		  &	0.51		       &      0.55		    &	      0.75		   &	9, 12$^{(\rm a)}$ \\		     
Mrk\,3$^{(\rm b, c)}$     &  4.84  		          &  0.23			 &    11.22		&		       &       ---		 & 2.85 	     &    3.21  		  &		 \multicolumn{2}{c}{1.45}		    &	      2.59		   &   10		  \\		     
ESO\,138\,G1$^{(\rm c)}$  &  $2.35\pm0.05$	          &  $0.34\pm0.02$		 &    $8.71\pm0.25$  	&		       &       ---		 & $3.01\pm0.10$     &    $0.68\pm0.03$ 	  &    $0.47\pm0.03$	       &      $0.48\pm0.03$	    &	      0.69		   &   11		  \\		     
55978-0990	          &  $6.76\pm 0.43 $        	  & $0.14\pm 0.06$    	   	 & $ 10.70\pm 0.73 $    &		       &     ---		 & $2.86\pm 0.31$    &  $3.30\pm 0.36$       &  $1.07\pm 0.18$  	  &   $0.87\pm 0.16$		&	$2.06\pm  0.66$       &   15	     \\
56104-0966	          &  $6.31\pm 0.14 $        	  & $0.07\pm 0.02$    	   	 & $  8.51\pm 0.20 $    &		       &     ---		 & $2.86\pm 0.09$    &  $1.04\pm 0.04$       &  $0.62\pm 0.03$  	  &   $0.56\pm 0.03$	       &       $0.73\pm  0.12$        &   15		    \\
55181-0154	          &  $4.12\pm 0.31 $        	  & $0.06\pm 0.03$    	   	 & $  7.82\pm 0.28 $    &		       &  $0.15\pm 0.04$	 & $2.86\pm 0.21$    &  $2.05\pm 0.15$       &  $0.60\pm 0.06$  	  &   $0.61\pm 0.06$	       &       $2.02\pm  0.42$        &   15		    \\
56088-0473	          &  $2.07\pm 0.01 $        	  & $0.06\pm 0.01$    	   	 & $  7.70\pm 0.04 $    &		       &  $0.01\pm 0.01$	 & $2.86\pm 0.01$    &  $0.15\pm 0.01$       &  $0.12\pm 0.01$  	  &   $0.11\pm 0.01$	       &       $0.63\pm  0.02$        &   15		    \\
56034-0154	          &  $3.02\pm 0.32 $        	  & $0.14\pm 0.01$    	   	 & $ 11.88\pm 0.23 $    &		       &  $0.05\pm 0.03$	 & $2.86\pm 0.07$    &  $2.13\pm 0.05$       &  $0.88\pm 0.03$  	  &   $0.81\pm 0.03$	       &       $2.10\pm  0.14$        &   15		    \\
56067-0382	          &  $5.51\pm 1.33 $        	  & $0.21\pm 0.10$    	   	 & $  8.70\pm 0.83 $    &   $0.192\pm0.165$       &	---		     & $2.86\pm 0.52$	 &  $2.36\pm 0.44$	 &  $0.86\pm 0.22$	      &   $0.81\pm 0.22$	   &	   $1.19\pm  0.89$	  &   15		\\
56240-0340	          &  $8.32\pm 0.52 $        	  & $0.16\pm 0.06$    	   	 & $  7.24\pm 0.56 $    &		       &    --- 		 & $2.86\pm 0.28$    &  $2.60\pm 0.26$       &  $1.10\pm 0.21$  	  &   $0.86\pm 0.20$	       &       $3.04\pm  0.94$        &   15		    \\
55539-0167	          &  $2.79\pm 0.20 $        	  & $0.05\pm 0.02$    	   	 & $  4.11\pm 0.08 $    &   $0.078\pm0.047$    &	---		     & $2.86\pm 0.08$	 &  $1.91\pm 0.05$	 &  $0.61\pm 0.03$	      &   $0.54\pm 0.03$	   &	   $1.08\pm  0.14$	  &   15		\\
56626-0636	          &  $3.82\pm 0.09 $        	  & $0.06\pm 0.01$    	   	 & $  6.12\pm 0.10 $    &		       &  $0.10\pm 0.02$	 & $2.86\pm 0.07$    &  $2.23\pm 0.05$       &  $1.06\pm 0.03$  	  &   $0.98\pm 0.03$	       &       $1.22\pm  0.10$        &   15		    \\
56074-0045	          &  $4.23\pm 0.32 $        	  & $0.03\pm 0.01$    	   	 & $  2.69\pm 0.04 $    &		       &    --- 		 & $2.86\pm 0.07$    &  $1.52\pm 0.04$       &  $0.77\pm 0.02$  	  &   $0.58\pm 0.02$	       &       $0.91\pm  0.11$        &   15		    \\
55212-0380	          &  $3.52\pm 1.13 $        	  & $0.10\pm 0.03$    	   	 & $ 12.54\pm 0.51 $    &		       &    --- 		 & $2.86\pm 0.15$    &  $2.94\pm 0.15$       &  $0.85\pm 0.06$  	  &   $0.84\pm 0.06$	       &       $2.41\pm  0.30$        &   15		    \\
56399-0302	          &  $5.01\pm 0.37 $        	  & $0.08\pm 0.04$    	   	 & $  3.07\pm 0.19 $    &		       &    --- 		 & $2.86\pm 0.23$    &  $2.16\pm 0.19$       &  $1.58\pm 0.16$  	  &   $1.14\pm 0.14$	       &       $1.33\pm  0.63$        &   15		    \\ 
56001-0293	          &  $5.05\pm 0.20 $        	  & $0.07\pm 0.01$    	   	 & $  7.63\pm 0.20 $    &   $0.047\pm0.037$    &    --- 		 & $2.86\pm 0.13$    &  $1.74\pm 0.08$       &  $0.85\pm 0.05$  	  &   $0.61\pm 0.04$	       &       $1.54\pm  0.21$        &   15		    \\ 
55651-0052	          &  $3.28\pm 0.76 $        	  & $0.02\pm 0.01$    	   	 & $  3.54\pm 0.04 $    &		       &  $0.05\pm 0.01$	 & $2.86\pm 0.06$    &  $2.06\pm 0.04$       &  $0.72\pm 0.02$  	  &   $0.68\pm 0.02$	       &       $0.94\pm  0.08$        &   15		    \\ 
56566-0794	          &  $6.58\pm 0.61 $        	  & $0.05\pm 0.02$    	   	 & $  4.74\pm 0.12 $    &   $0.017\pm0.016$    &  $0.08\pm 0.03$	 & $2.86\pm 0.10$    &  $2.16\pm 0.08$       &  $1.26\pm 0.06$  	  &   $1.07\pm 0.05$	       &       $1.22\pm  0.16$        &   15		    \\ 
56748-0336	          &  $4.97\pm 0.51 $        	  & $0.08\pm 0.05$    	   	 & $  7.11\pm 0.41 $    &		       &    --- 		 & $2.86\pm 0.24$    &  $3.23\pm 0.27$       &  $0.99\pm 0.12$  	  &   $0.89\pm 0.12$	       &       $1.85\pm  0.50$        &   15		    \\ 
56206-0454	          &  $7.35\pm 0.27 $        	  & $0.09\pm 0.02$    	   	 & $  6.67\pm 0.19 $    &		       &  $0.05\pm 0.03$	 & $2.86\pm 0.11$    &  $1.21\pm 0.06$       &  $1.12\pm 0.05$  	  &   $0.91\pm 0.05$	       &       $1.08\pm  0.19$        &   15		    \\ 
55860-0112	          &  $5.74\pm 1.44 $        	  & $0.11\pm 0.07$    	   	 & $  7.82\pm 0.66 $    &		       &  $0.15\pm 0.12$	 & $2.86\pm 0.33$    &  $5.38\pm 0.61$       &  $1.80\pm 0.23$  	  &   $1.77\pm 0.23$	       &       $1.78\pm  0.46$        &   15		    \\ 
56453-0144	          &  $3.19\pm 0.37 $        	  & $0.06\pm 0.02$    	   	 & $  6.42\pm 0.18 $    &		       &    --- 		 & $2.86\pm 0.10$    &  $2.64\pm 0.10$       &  $0.89\pm 0.06$  	  &   $0.78\pm 0.06$	       &       $2.20\pm  0.31$        &   15		    \\ 
55710-0116	          &  $2.72\pm 1.05 $        	  & $0.07\pm 0.02$    	   	 & $  7.54\pm 0.24 $    &		       &  $0.06\pm 0.04$	 & $2.86\pm 0.12$    &  $4.47\pm 0.18$       &  $1.04\pm 0.06$  	  &   $1.13\pm 0.07$	       &       $2.58\pm  0.21$        &   15		    \\ 
56209-0390	          &  $6.99\pm 0.77 $        	  & $0.07\pm 0.06$    	   	 & $  8.57\pm 0.97 $    &		       &    --- 		 & $2.86\pm 0.58$    &  $2.75\pm 0.55$       &  $0.39\pm 0.15$  	  &   $0.22\pm 0.14$	       &       $0.52\pm  0.65$        &   15		    \\ 
56298-0302	          &  $10.16\pm 1.84$        	  & $0.09\pm 0.07$    	   	 & $  5.39\pm 0.46 $    &		       &    --- 		 & $2.86\pm 0.42$    &  $3.29\pm 0.49$       &  $1.59\pm 0.27$  	  &   $1.40\pm 0.24$	       &       $7.00\pm  2.26$        &   15		    \\ 
56366-0928	          &  $4.81\pm 0.41 $        	  & $0.12\pm 0.05$    	   	 & $  5.70\pm 0.40 $    &		       &  $0.21\pm 0.08$	 & $2.86\pm 0.28$    &  $2.51\pm 0.25$       &  $0.96\pm 0.12$  	  &   $0.96\pm 0.12$	       &       $2.31\pm  0.56$        &   15		    \\ 
55742-0383	          &  $2.74\pm 0.55 $        	  & $0.05\pm 0.01$    	   	 & $  4.99\pm 0.10 $    &   $0.049\pm0.038$    &    --- 		 & $2.86\pm 0.08$    &  $2.06\pm 0.06$       &  $0.63\pm 0.03$  	  &   $0.56\pm 0.03$	       &       $1.01\pm  0.14$        &   15		    \\
55302-0655	          &  $4.90\pm 0.18 $        	  & $0.05\pm 0.02$    	   	 & $  2.06\pm 0.05 $    &   $0.035\pm0.019$    &	---		     & $2.86\pm 0.08$	 &  $1.67\pm 0.06$	 &  $1.24\pm 0.04$	      &   $1.00\pm 0.04$	   &	   $0.94\pm  0.14$	  &   15		\\
56328-0550	          &  $6.45\pm 0.26 $        	  & $0.06\pm 0.02$    	   	 & $  4.90\pm 0.15 $    &		       &  $0.06\pm 0.04$	 & $2.86\pm 0.13$    &  $3.35\pm 0.16$       &  $1.10\pm 0.06$  	  &   $0.86\pm 0.06$	       &       $1.57\pm  0.25$        &   15		    \\
55646-0770	          &  $8.48\pm 2.32 $        	  & $0.14\pm 0.05$    	   	 & $  7.32\pm 1.31 $    &		       &  ---			 & $2.86\pm 0.95$    &  $3.74\pm 1.23$       &  $1.50\pm 0.56$  	  &   $1.39\pm 0.52$	       &       $1.82\pm  1.49$        &   15		    \\
55617-0758	          &  $8.05\pm 0.33 $        	  & $0.05\pm 0.04$    	   	 & $  3.84\pm 0.20 $    &		       &  $0.11\pm 0.06$	 & $2.86\pm 0.27$    &  $2.26\pm 0.22$       &  $1.24\pm 0.14$  	  &   $1.38\pm 0.15$	       &       $1.33\pm  0.40$        &   15		    \\
56003-0218	          &  $3.35\pm 1.00 $        	  & $0.07\pm 0.04$    	   	 & $  4.70\pm 0.29 $    &		       &  $0.10\pm 0.07$	 & $2.86\pm 0.22$    &  $1.68\pm 0.15$       &  $0.56\pm 0.09$  	  &   $0.58\pm 0.09$	       &       $2.06\pm  0.45$        &   15		    \\
55629-0364	          &  $9.11\pm 0.64 $        	  & $0.11\pm 0.07$    	   	 & $  7.08\pm 0.52 $    &		       &    --- 		 & $2.86\pm 0.41$    &  $3.91\pm 0.57$       &  $1.44\pm 0.24$  	  &   $1.32\pm 0.23$	       &       $1.64\pm  0.80$        &   15		    \\
56568-0076	          &  $5.12\pm 0.59 $        	  & $0.03\pm 0.01$    	   	 & $  3.16\pm 0.05 $    &   $0.031\pm0.021$    &	---		     & $2.86\pm 0.09$	 &  $2.98\pm 0.10$	 &  $0.71\pm 0.03$	      &   $0.69\pm 0.03$	   &	   $0.66\pm  0.09$	  &   15		\\
55836-0160	          &  $3.57\pm 0.33 $        	  & $0.04\pm 0.02$    	   	 & $  3.70\pm 0.11 $    &		       &    --- 		 & $2.86\pm 0.14$    &  $2.24\pm 0.11$       &  $0.67\pm 0.05$  	  &   $0.59\pm 0.05$	       &       $1.71\pm  0.32$        &   15		    \\
55505-0654	          &  $3.05\pm 0.19 $        	  & $0.04\pm 0.02$    	   	 & $  7.95\pm 0.28 $    &		       &  $0.05\pm 0.03$	 & $2.86\pm 0.17$    &  $2.70\pm 0.16$       &  $0.65\pm 0.05$  	  &   $0.58\pm 0.05$	       &       $2.10\pm  0.35$        &   15		    \\
\hline												
\end{tabular}
\begin{minipage}[c]{2\columnwidth}
References--- (1) \citet{2015ApJS..217...12D}, (2) \citet{1986A&A...166...92B}, (3) \citet{1983ApJ...266..485P}, (4) \citet{1978ApJ...223...56K}, (5) \citet{1994ApJ...435..171K}, (6) \citet{1980ApJ...240...32S},
(7) \citet{2017ApJS..232...11T}, (8) \citet{1983ApJ...273..489C}, (6) \citet{1983ApJ...269..416G}, (10) \citet{1983ApJ...265...92M}, (11) \citet{1992A&A...266..117A}.
(12) \citet{2006A&A...457...61R}, (13) \citet{2011ApJ...743..100R}, (14) \citet{2006ApJ...645..148R}, (15) BOSS/SDSS sample,
(a) Reference which  the [\ion{S}{iii}]$\lambda9069$ emission line fluxes  was compiled.
(b) Emission-line intensities
corrected by reddening in the present work. (c) Object which the [\ion{S}{iii}]$\lambda9069$ was measured with the others. 
\end{minipage}
\end{table*}

\begin{table*}
\caption{Electron density and electron temperature values assumed in the ionic abundance calculations. $T_{\rm e}(\rm O^{2+})$ was calculated
through the observational $RO3$ line ratio, whose line intensities are listed in Table~\ref{tab1}, and by using the \textsc{PyNeb} code \citep{Luridiana2015}.
$T_{\rm e}(\rm N^{+})$ and $T_{\rm e}(\rm S^{2+})$ were
calculated either from $RN2$ and $RS3$ line ratios (when these line ratios were measured), respectively, or from
Eq.~\ref{eq3} and \ref{eq4} (when  auroral lines were not measured, see Table~\ref{tab1}).}
\label{tab2}
\begin{tabular}{@{}lcccc@{}}	 
\noalign{\smallskip} 
\hline 
 Object      & $T_{\rm e}(\rm O^{2+})$ K & $T_{\rm e}(\rm S^{2+})$ K & $T_{\rm e}(\rm N^{+})$ K   & $N_{\rm e} (\rm cm^{-3}$)  \\
\hline
Mrk\,573     & $12888\pm 425$ & $21341\pm7881$  & $11673\pm1899$ & $ 559\pm102$ \\
NGC\,5728    & $12301\pm 725$ & $20030\pm7929$  & $11373\pm1923$ & $ 674\pm167$ \\
ESO\,428-G14 & $15651\pm1197$ & $27501\pm8055$  & $13082\pm1983$ & $1059\pm466$ \\
NGC\,4388    & $13446\pm1254$ & $22585\pm8074$  & $11958\pm1992$ &   393:       \\
Mrk\,78      & $12162\pm1139$ & $19721\pm8036$  & $11303\pm1974$ &   475:       \\
NGC\,7674    & $11296\pm1126$ & $17790\pm8032$  & $10861\pm1972$ & $1396\pm620$ \\
NGC\,2110    & $20761\pm3395$ & $38896\pm9338$  & $15688\pm2561$ & $ 691\pm300$ \\
NGC\,7682    & $14146\pm1412$ & $24146\pm8131$  & $12315\pm2020$ & $1067\pm495$ \\
NGC\,3227    & $25028\pm2500$ & $48412\pm17000$ & $17864\pm5800$ & $1067\pm485$ \\
Mrk\,1066    & $15446\pm3553$ & $27045\pm9466$  & $12978\pm2616$ &  1122:       \\
Mrk\,3       & $15335\pm1185$ & $26796\pm8051$  & $12921\pm1981$ &   500:       \\
ESO\,138 G1  & $21996\pm1318$ & $41650\pm8097$  & $16318\pm2003$ & $1039\pm452$ \\
55978-0990   & $12685\pm2490$ & $20888\pm8684$  & $11570\pm2274$ &   266:       \\
56104-0966   & $10754\pm1095$ & $16582\pm8023$  & $10585\pm1968$ & $ 471\pm210$ \\
55181-0154   & $10508\pm1855$ & $24392\pm9679$  & $10459\pm2111$ & $ 778\pm384$ \\
56088-0473   & $10566\pm 301$ & $ 9244\pm1351$  & $10488\pm1893$ & $ 502\pm186$ \\
56034-0154   & $12184\pm 409$ & $11052\pm2481$  & $11314\pm1898$ & $ 537\pm167$ \\
56067-0382   & $16612\pm5030$ & $29644\pm10845$ & $31422\pm9900$ &   666:       \\
56240-0340   & $15912\pm3588$ & $28083\pm9495$  & $13215\pm2628$ &   200:       \\
55539-0167   & $12334\pm1976$ & $20105\pm8387$  & $16670\pm5500$ & $ 445\pm218$ \\
56626-0636   & $11409\pm 726$ & $27152\pm5997$  & $10919\pm1923$ & $ 537\pm137$ \\
56074-0045   & $11954\pm1531$ & $19258\pm8179$  & $11197\pm2042$ & $ 127\pm58$  \\
55212-0380   & $10644\pm1185$ & $16335\pm8051$  & $10528\pm1981$ & $ 699\pm315$ \\
56399-0302   & $17311\pm5347$ & $31203\pm11175$ & $13929\pm3316$ &    50:       \\
56001-0293   & $11159\pm 640$ & $17484\pm7913$  & $13152\pm4500$ &    55:       \\
55651-0052   & $ 9576\pm 733$ & $18122\pm2797$  & $ 9984\pm1923$ & $ 555\pm130$ \\
56566-0794   & $11713\pm1806$ & $21676\pm7539$  & $10771\pm3800$ & $ 347\pm162$ \\
56748-0336   & $11981\pm3128$ & $19319\pm9129$  & $11211\pm2471$ &   477:       \\
56206-0454   & $12842\pm1266$ & $16315\pm7012$  & $11649\pm1994$ & $ 266\pm130$ \\
55860-0112   & $13038\pm3931$ & $27975\pm12880$ & $11749\pm2753$ & $ 735\pm360$ \\
56453-0144   & $11224\pm1400$ & $17630\pm8127$  & $10824\pm2017$ & $ 410\pm202$ \\
55710-0116   & $11189\pm1218$ & $10917\pm2756$  & $10806\pm1986$ & $1021\pm381$ \\
56209-0390   & $10735\pm3862$ & $16538\pm8150$  & $10575\pm2728$ & $ 198\pm49$  \\
56298-0302   & $14006\pm5604$ & $23833\pm11451$ & $12243\pm3425$ & $ 451\pm215$ \\
56366-0928   & $15523\pm3682$ & $30648\pm9727$  & $13017\pm2662$ & $ 839\pm410$ \\
55742-0383   & $11497\pm 889$ & $18239\pm7966$  & $12245\pm3500$ & $ 445\pm206$ \\
55302-0655   & $16672\pm3674$ & $29778\pm9568$  & $11547\pm3178$ & $ 267\pm122$ \\
56328-0550   & $12373\pm1714$ & $14346\pm5621$  & $11410\pm2080$ &   192:       \\
55646-0770   & $14856\pm3805$ & $25729\pm9680$  & $12676\pm2706$ &   593:       \\
55617-0758   & $12640\pm4540$ & $27347\pm10109$ & $11546\pm2987$ & $1168\pm560$ \\
56003-0218   & $13337\pm3563$ & $16884\pm8330$  & $11902\pm2620$ & $ 913\pm430$ \\
55629-0364   & $13575\pm4192$ & $22872\pm10028$ & $12023\pm2852$ &   546:       \\
56568-0076   & $11281\pm1368$ & $17756\pm8115$  & $10853\pm2012$ & $ 666\pm215$ \\
55836-0160   & $11814\pm2314$ & $18946\pm8576$  & $11125\pm2225$ & $ 427\pm205$ \\
55505-0654   & $ 9268\pm1446$ & $11052\pm2955$  & $ 9827\pm2026$ & $ 422\pm190$ \\
\hline
\multicolumn{5}{l}{Note. For some objects we use ``:'' to indicate that error bars are at least an order} \\
\multicolumn{5}{l}{of magnitude larger than the expected density. This is due to the significant} \\
\multicolumn{5}{l}{emission-line errors (see Sect~\ref{sectene}).}
\end{tabular}	   									
\end{table*}	

\begin{table*}
\caption{Ionic and total abundances for Seyfert~2 nuclei derived through the $T_{\rm e}$-method.}
\label{tab3}
\begin{tabular}{@{}lcccccccc@{}}	 
\noalign{\smallskip} 
\hline 
 Object         &  $\rm O^{+}/H^{+}$   &	$\rm O^{2+}/H^{+}$  & O/H       &  $\rm S^{+}/H^{+}$ &  $\rm S^{2+}/H^{+}$ & ICF(S)   & S/H & S/O         \\ 
\hline  
Mrk\,573     & $7.95\pm0.12$ & $8.21\pm0.04$ & $8.58\pm0.13$ & $6.46\pm0.12$ & $6.52\pm0.16$ & $2.02$ & $7.08\pm0.18$ & $-1.50\pm0.22$ \\
NGC\,5728    & $8.03\pm0.12$ & $8.22\pm0.07$ & $8.61\pm0.14$ & $6.52\pm0.12$ & $6.42\pm0.17$ & $1.81$ & $7.07\pm0.18$ & $-1.55\pm0.23$ \\
ESO\,428-G14 & $7.79\pm0.10$ & $8.03\pm0.07$ & $8.40\pm0.12$ & $6.44\pm0.10$ & $6.24\pm0.15$ & $1.95$ & $6.94\pm0.16$ & $-1.46\pm0.20$ \\
NGC\,4388    & $8.03\pm0.12$ & $8.17\pm0.09$ & $8.59\pm0.13$ & $6.37\pm0.12$ & $6.12\pm0.16$ & $1.83$ & $6.85\pm0.18$ & $-1.74\pm0.22$ \\
Mrk\,78      & $8.38\pm0.13$ & $8.35\pm0.10$ & $8.84\pm0.14$ & $6.41\pm0.13$ & $6.10\pm0.17$ & $1.52$ & $6.87\pm0.19$ & $-1.97\pm0.23$ \\
NGC\,7674    & $7.85\pm0.13$ & $8.47\pm0.10$ & $8.74\pm0.15$ & $6.47\pm0.13$ & $6.71\pm0.18$ & $2.94$ & $7.20\pm0.19$ & $-1.54\pm0.24$ \\
NGC\,2110    & $7.84\pm0.10$ & $7.39\pm0.09$ & $8.15\pm0.11$ & $6.50\pm0.10$ & $6.30\pm0.14$ & $1.22$ & $6.71\pm0.16$ & $-1.43\pm0.19$ \\
NGC\,7682    & $7.99\pm0.12$ & $8.06\pm0.09$ & $8.50\pm0.13$ & $6.61\pm0.11$ & $6.33\pm0.16$ & $1.61$ & $7.08\pm0.17$ & $-1.42\pm0.22$ \\
NGC\,3227    & $7.55\pm0.14$ & $7.59\pm0.06$ & $8.04\pm0.15$ & $6.36\pm0.15$ & $6.28\pm0.19$ & $1.72$ & $6.62\pm0.21$ & $-1.42\pm0.26$ \\
Mrk\,1066    & $7.98\pm0.13$ & $7.58\pm0.11$ & $8.30\pm0.14$ & $6.24\pm0.13$ & $5.85\pm0.17$ & $1.21$ & $6.68\pm0.19$ & $-1.62\pm0.24$ \\
Mrk\,3       & $8.16\pm0.11$ & $8.05\pm0.07$ & $8.58\pm0.12$ & $6.62\pm0.11$ & $6.40\pm0.15$ & $1.46$ & $7.11\pm0.17$ & $-1.47\pm0.21$ \\
ESO\,138 G1  & $7.52\pm0.08$ & $7.60\pm0.05$ & $8.04\pm0.09$ & $6.00\pm0.08$ & $5.78\pm0.12$ & $1.74$ & $6.20\pm0.14$ & $-1.83\pm0.16$ \\
55978-0990   & $8.49\pm0.14$ & $8.25\pm0.09$ & $8.86\pm0.15$ & $6.55\pm0.14$ & $6.43\pm0.18$ & $1.35$ & $7.09\pm0.20$ & $-1.78\pm0.25$ \\
56104-0966   & $8.60\pm0.14$ & $8.37\pm0.11$ & $8.97\pm0.15$ & $6.44\pm0.14$ & $6.12\pm0.18$ & $1.32$ & $6.90\pm0.20$ & $-2.08\pm0.25$ \\
55181-0154   & $8.42\pm0.15$ & $8.36\pm0.11$ & $8.87\pm0.16$ & $6.48\pm0.15$ & $6.34\pm0.19$ & $1.43$ & $7.01\pm0.21$ & $-1.86\pm0.26$ \\
56088-0473   & $8.13\pm0.14$ & $8.35\pm0.04$ & $8.73\pm0.15$ & $5.74\pm0.14$ & $6.50\pm0.18$ & $1.89$ & $6.86\pm0.20$ & $-1.87\pm0.25$ \\
56034-0154   & $8.16\pm0.12$ & $8.35\pm0.04$ & $8.74\pm0.14$ & $6.53\pm0.12$ & $6.87\pm0.17$ & $1.87$ & $7.32\pm0.18$ & $-1.41\pm0.23$ \\
56067-0382   & $7.58\pm0.16$ & $7.86\pm0.13$ & $8.04\pm0.17$ & $6.03\pm0.15$ & $6.01\pm0.19$ & $1.31$ & $6.32\pm0.21$ & $-1.72\pm0.27$ \\
56240-0340   & $8.38\pm0.13$ & $7.82\pm0.10$ & $8.66\pm0.14$ & $6.43\pm0.13$ & $6.44\pm0.17$ & $1.18$ & $7.03\pm0.19$ & $-1.63\pm0.23$ \\
55539-0167   & $7.58\pm0.15$ & $7.87\pm0.10$ & $8.22\pm0.16$ & $6.03\pm0.15$ & $6.17\pm0.20$ & $1.32$ & $6.70\pm0.22$ & $-1.53\pm0.27$ \\
56626-0636   & $8.32\pm0.13$ & $8.14\pm0.07$ & $8.72\pm0.14$ & $6.65\pm0.13$ & $6.06\pm0.17$ & $1.35$ & $7.04\pm0.19$ & $-1.68\pm0.24$ \\
56074-0045   & $8.35\pm0.13$ & $7.73\pm0.09$ & $8.62\pm0.14$ & $6.41\pm0.13$ & $6.12\pm0.17$ & $1.15$ & $6.88\pm0.19$ & $-1.74\pm0.24$ \\
55212-0380   & $8.34\pm0.14$ & $8.55\pm0.10$ & $8.94\pm0.15$ & $6.62\pm0.14$ & $6.65\pm0.18$ & $1.81$ & $7.22\pm0.20$ & $-1.71\pm0.25$ \\
56399-0302   & $8.10\pm0.14$ & $7.37\pm0.12$ & $8.35\pm0.15$ & $6.51\pm0.14$ & $6.47\pm0.18$ & $1.13$ & $6.79\pm0.20$ & $-1.56\pm0.25$ \\
56001-0293   & $8.19\pm0.17$ & $8.27\pm0.07$ & $8.71\pm0.18$ & $6.29\pm0.15$ & $6.41\pm0.19$ & $1.38$ & $6.95\pm0.21$ & $-1.76\pm0.28$ \\
55651-0052   & $8.41\pm0.15$ & $8.16\pm0.10$ & $8.78\pm0.16$ & $6.58\pm0.15$ & $6.17\pm0.19$ & $1.28$ & $7.01\pm0.21$ & $-1.77\pm0.26$ \\
56566-0794   & $8.57\pm0.13$ & $8.00\pm0.11$ & $8.85\pm0.14$ & $7.06\pm0.16$ & $6.18\pm0.20$ & $1.16$ & $7.40\pm0.22$ & $-1.45\pm0.26$ \\
56748-0336   & $8.39\pm0.15$ & $8.14\pm0.12$ & $8.76\pm0.16$ & $6.58\pm0.15$ & $6.43\pm0.19$ & $1.32$ & $7.10\pm0.21$ & $-1.66\pm0.26$ \\
56206-0454   & $8.51\pm0.12$ & $8.03\pm0.10$ & $8.81\pm0.14$ & $6.56\pm0.12$ & $6.30\pm0.17$ & $1.20$ & $7.04\pm0.18$ & $-1.77\pm0.23$ \\
55860-0112   & $8.37\pm0.15$ & $8.08\pm0.12$ & $8.72\pm0.16$ & $6.83\pm0.15$ & $6.21\pm0.19$ & $1.27$ & $7.22\pm0.21$ & $-1.51\pm0.27$ \\
56453-0144   & $8.26\pm0.14$ & $8.19\pm0.09$ & $8.70\pm0.15$ & $6.56\pm0.14$ & $6.56\pm0.18$ & $1.47$ & $7.15\pm0.20$ & $-1.55\pm0.25$ \\
55710-0116   & $8.18\pm0.13$ & $8.26\pm0.11$ & $8.70\pm0.15$ & $6.72\pm0.13$ & $6.97\pm0.18$ & $1.58$ & $7.45\pm0.20$ & $-1.25\pm0.24$ \\
56209-0390   & $8.66\pm0.17$ & $8.38\pm0.13$ & $9.02\pm0.18$ & $6.12\pm0.14$ & $5.97\pm0.19$ & $1.32$ & $6.65\pm0.20$ & $-2.37\pm0.27$ \\
56298-0302   & $8.56\pm0.16$ & $7.83\pm0.13$ & $8.81\pm0.17$ & $6.70\pm0.15$ & $6.89\pm0.19$ & $1.10$ & $7.40\pm0.21$ & $-1.42\pm0.27$ \\
56366-0928   & $8.13\pm0.13$ & $7.74\pm0.11$ & $8.46\pm0.14$ & $6.48\pm0.13$ & $6.81\pm0.17$ & $1.22$ & $6.98\pm0.19$ & $-1.48\pm0.24$ \\
55742-0383   & $7.99\pm0.16$ & $8.04\pm0.09$ & $8.50\pm0.17$ & $6.30\pm0.15$ & $6.20\pm0.20$ & $1.41$ & $6.84\pm0.21$ & $-1.65\pm0.28$ \\
55302-0655   & $8.35\pm0.16$ & $7.23\pm0.11$ & $8.56\pm0.18$ & $6.61\pm0.15$ & $5.91\pm0.20$ & $1.09$ & $6.98\pm0.21$ & $-1.58\pm0.28$ \\
56328-0550   & $8.50\pm0.13$ & $7.94\pm0.11$ & $8.78\pm0.14$ & $6.56\pm0.13$ & $6.55\pm0.17$ & $1.17$ & $7.15\pm0.19$ & $-1.63\pm0.24$ \\
55646-0770   & $8.42\pm0.13$ & $7.90\pm0.11$ & $8.71\pm0.15$ & $6.66\pm0.14$ & $6.26\pm0.18$ & $1.17$ & $7.10\pm0.20$ & $-1.62\pm0.25$ \\
55617-0758   & $8.54\pm0.16$ & $7.81\pm0.12$ & $8.79\pm0.17$ & $6.74\pm0.16$ & $6.10\pm0.21$ & $1.09$ & $7.12\pm0.22$ & $-1.67\pm0.28$ \\
56003-0218   & $8.11\pm0.14$ & $7.83\pm0.11$ & $8.47\pm0.15$ & $6.34\pm0.14$ & $6.56\pm0.19$ & $1.27$ & $7.05\pm0.20$ & $-1.42\pm0.26$ \\
55629-0364   & $8.54\pm0.15$ & $7.99\pm0.12$ & $8.82\pm0.16$ & $6.69\pm0.15$ & $6.28\pm0.19$ & $1.16$ & $7.12\pm0.21$ & $-1.70\pm0.26$ \\
56568-0076   & $8.45\pm0.14$ & $7.87\pm0.10$ & $8.73\pm0.15$ & $6.50\pm0.13$ & $6.03\pm0.18$ & $1.13$ & $6.92\pm0.20$ & $-1.81\pm0.25$ \\
55836-0160   & $8.26\pm0.14$ & $7.88\pm0.11$ & $8.59\pm0.15$ & $6.41\pm0.14$ & $6.41\pm0.18$ & $1.23$ & $7.00\pm0.20$ & $-1.59\pm0.25$ \\
55505-0654   & $8.42\pm0.15$ & $8.56\pm0.11$ & $8.97\pm0.17$ & $6.53\pm0.16$ & $6.87\pm0.20$ & $1.74$ & $7.33\pm0.22$ & $-1.65\pm0.27$ \\
\hline  	   											
\end{tabular}	   									
\end{table*}

\bsp	
\label{lastpage}
\end{document}